%% file: valence.tex
\documentclass{article} % For LaTeX2e
\usepackage{valence,spectral}

\input{math_commands.tex}

\usepackage{hyperref}
\usepackage{url}
\usepackage{graphicx}
\usepackage{xcolor}
\definecolor{grey}{RGB}{140, 146, 172}
\definecolor{darkred}{RGB}{184, 12, 43}

\hypersetup{
    colorlinks = true,
    linkcolor=darkred,
    urlcolor=darkred,
    citecolor=grey,
} 

\usepackage{makecell} % Include makecell package to use \thead

\usepackage[utf8]{inputenc} % allow utf-8 input
\usepackage[T1]{fontenc}    % use 8-bit T1 fonts
\usepackage{url}            % simple URL typesetting
\usepackage{booktabs}       % professional-quality tables
\usepackage{amsfonts}       % blackboard math symbols
\usepackage{nicefrac}       % compact symbols for 1/2, etc.
\usepackage{microtype}      % microtypography
\usepackage{amsmath,amssymb}
\usepackage{graphicx}
\usepackage{mwe}
\usepackage{subcaption}
\usepackage{float}
\usepackage{tikz}
\usepackage{pifont}%
\usepackage{adjustbox}
\usepackage{wrapfig}
\usepackage{enumitem}
\usepackage[normalem]{ulem}
\usepackage{stackengine}
\usepackage{mathtools}
\usepackage{tcolorbox}

\newcommand{\cmark}{\ding{51}}%
\newcommand{\xmark}{\ding{55}}%
\definecolor{babyblue}{rgb}{0.54, 0.81, 0.94}
\definecolor{dark2orange}{RGB}{237, 127, 49}

\newtheorem{theorem}{Guideline}

\title{How Molecules Impact Cells: \\Unlocking Contrastive PhenoMolecular Retrieval}

\author{%
  Philip Fradkin\(^{1,3,*}\),
  Puria Azadi$^{1, 2, } \thanks{These two authors contributed equally and reserve the right to swap their order.} \:\:$,
  \textbf{Karush Suri}$^{1}$,\\
  \textbf{Frederik Wenkel}$^{1}$, \textbf{Ali Bashashati}$^{2}$ \textbf{Maciej Sypetkowski}$^{1}\thanks{Equal advising.} \;$, \textbf{Dominique Beaini}$^{1,4, \dagger}$\\
  $^{1}$ Valence Labs, $^{2}$ University of British Columbia,\\
  $^{3}$ University of Toronto, Vector Institute, $^{4}$ Universit\'{e} de Montr\'{e}al, \\ Mila- Quebec AI Institute\\
  \texttt{dominique@valencelabs.com}
}

\iclrfinalcopy
\begin{document}

\maketitle

\begin{abstract}

\input{abstract}
\end{abstract}

\tableofcontents

\input{introduction}

\input{related_work}

\input{methodology}

\input{experiments}

\input{conclusion}

\newpage

\bibliography{refs}
\bibliographystyle{valence}

\input{appendix}

\end{document}

%% file: math_commands.tex
\usepackage{amsmath,amsfonts,bm}

\def\eqref#1{equation~\ref{#1}}
\def\ceil#1{\lceil #1 \rceil}

\def\1{\bm{1}}

\DeclareMathAlphabet{\mathsfit}{\encodingdefault}{\sfdefault}{m}{sl}
\SetMathAlphabet{\mathsfit}{bold}{\encodingdefault}{\sfdefault}{bx}{n}

%% file: abstract.tex
% Introduce small molecules impact on cells
Predicting molecular impact on cellular function is a core challenge in therapeutic design. 
Phenomic experiments, designed to capture cellular morphology, utilize microscopy based techniques and demonstrate a high throughput solution for uncovering molecular impact on the cell. 
In this work, we learn a joint latent space between molecular structures and microscopy phenomic experiments, aligning paired samples with contrastive learning. 
Specifically, we study the problem of \textit{Contrastive PhenoMolecular Retrieval}, which consists of zero-shot molecular structure identification conditioned on phenomic experiments. 
We assess challenges in multi-modal learning of phenomics and molecular modalities such as experimental batch effect, inactive molecule perturbations, and encoding perturbation concentration.
We demonstrate improved multi-modal learner retrieval through (1) a uni-modal pre-trained phenomics model, (2) a novel inter sample similarity aware loss, and (3) models conditioned on a representation of molecular concentration.
Following this recipe, we propose \textit{MolPhenix}, a \underline{mol}ecular \underline{phen}om\underline{ics} model.
MolPhenix leverages a pre-trained phenomics model to demonstrate significant performance gains across perturbation concentrations, molecular scaffolds, and activity thresholds. 
% Workshop this
In particular, we demonstrate an 8.1$\times$ improvement in zero shot molecular retrieval of active molecules over the previous state-of-the-art, reaching 77.33\% in top-1\% accuracy. 
These results open the door for machine learning to be applied in virtual phenomics screening, which can significantly benefit drug discovery applications.
%These results open the door for machine learning usage in virtual phenomics screening of molecular perturbations, and show how foundational models of biology and chemistry can play an essential role in drug discovery. 

%% file: introduction.tex
\section{Introduction}

Quantifying cellular responses elicited by genetic and molecular perturbations represents a core challenge in medicinal research \cite{crispr_cell_impact, cell_impact2}. 
Out of an approximate \(10^{60}\) druglike molecule designs, a small number are able to alter cellular properties to reverse the course of diseases \cite{perturbation_space,drugbank}. 
In recent years, microscopy-based cell morphology screening techniques, demonstrated potential for quantitative understanding of a molecule's biological effects. 
Experimental techniques such as cell-painting are used to capture cellular morphology, which correspond to physical and structural properties of the cell \cite{image-profiling, cell_painting}. 
Cells treated with molecular perturbations can change morphology, which is captured by staining and high throughput microscopy techniques. 
Perturbations with similar cellular impact induce analogous morphological changes, allowing to capture underlying biological effects in phenomic experiments. Identifying such perturbations with similar morphological changes can aid in discovery of novel therapeutic drug candidates \cite{extracted_feature_method, HCS-deep-learning, HCS-deep-learning2}.

% Introduce mulit-modal learning
Determining molecular impact on the cell can be formulated as a multi-modal learning problem, allowing us to build on a rich family of methods \cite{clip,siglip,cwcl}. Similar to text-image models, paired data is collected from phenomic experiments along with molecules used to perturb the cells. Contrastive objectives have been used as an effective approach in aligning paired samples from different modalities \cite{clip, lanusse2023astroclip}. A model that has learned a cross-modal joint latent space must be able to retrieve a molecular perturbant conditioned on the phenomic experiment. We identify this problem as \textit{contrastive phenomolecular retrieval} (see Figure \ref{fig:contrastive_phenomolecular_retrieval}). 
% Do we need a new paragraph ere?
Addressing this problem can allow for identification of molecular impact on cellular function, however, this comes with its own set of challenges. 
 \cite{cloob, surv_contrastive, better_features_contrastive}. 
% Molecular information is retrieved from a joint embedding space of phenomic microscopy images and molecule fingerprints constructed using multi-modal contrastive learning. 
% Our work, thus, identifies the three main challenges affecting contrastive phenomolecular retrieval.

% Low data and batch effect affecting generalization
\textbf{(1)} Firstly, multi-modal paired phenomics molecular data suffers from lower overall dataset sizes and is subject to batch effects. Challenges with uniform processing and prohibitive costs associated with acquisition of paired data, leads to an order of magnitude fewer data points compared to text-image datasets \cite{laion,jump_cp}. Furthermore, data is subject to random batch effects that capture non-biologically meaningful variation \cite{lek_batch_effect,rxrx1}. 
% Inactive molecules
\textbf{(2)} Paired phenomic-molecular data contains inactive perturbations that do not have a biological effect or do not perturb cellular morphology. It is difficult to infer a priori whether a molecule has a cellular effect, leading to the collection of paired molecular structures with unperturbed cells. These data-points are challenging to filter out without an effective phenomic embedding, as morphological effects are rarely discernible. These samples can be interpreted as misannotated, under the assumption of all collected pairs having biologically meaningful interactions.
% Effects of dosage on the molecule
\textbf{(3)} Finally, a complete solution for capturing molecular effects on cells must capture molecular concentration. The same molecule can have drastically different effects along its dose response curve, thus making concentration an essential component for learning molecular impact.

In this work, we explore the problem of contrastive phenomolecular retrieval by addressing the above challenges circumvented in prior works. Our key contributions are as follows:

\begin{itemize}[leftmargin=*]
    \item We demonstrate significantly higher phenomolecular retrieval rates by utilizing a pretrained uni-modal phenomic encoder. Thus alleviating the data availability challenge, reducing the impact of batch effects, and identifying molecular activity levels. 
    \item We propose a novel soft-weighted sigmoid locked loss (\texttt{S2L}) that addresses the effects of inactive molecules. This is done by leveraging distances computed in the phenomic embedding space to learn inter-sample similarities.
    \item We explore \textit{explicit} and \textit{implicit} methods to encode molecular concentration, assessing the model's ability to perform retrieval in an inter-concentration setting and generalize to  unseen concentrations.
\end{itemize}

\begin{figure}[H]
    \centering
    \includegraphics[width=\textwidth]{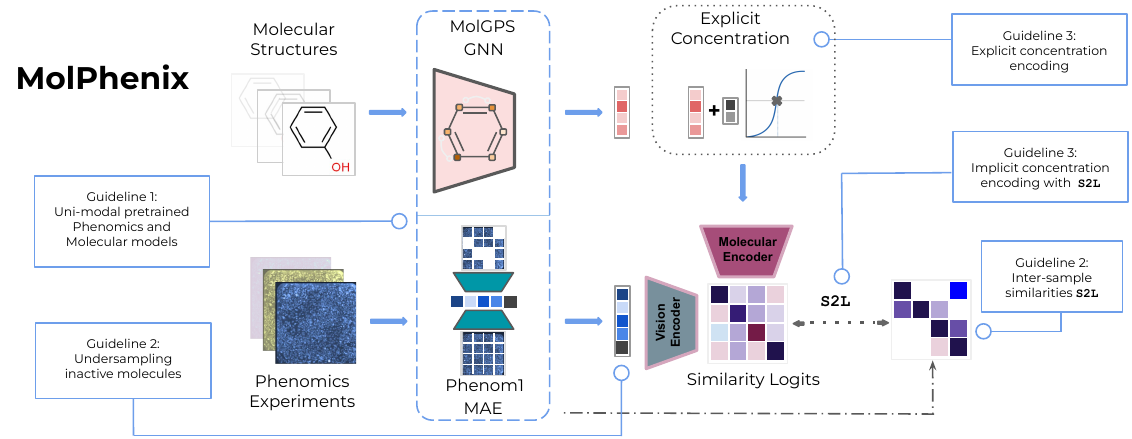}
    \caption{Illustration of proposed guidelines when incorporated in our \textit{MolPhenix} contrastive phenomolecular retrieval framework. We address challenges by utilizing uni-modal pretrained MAE \& MPNN models, inter-sample weighting with a dosage aware \texttt{S2L} loss, undersampling inactive molecules, and encoding molecular concentration.}
    \label{fig:main}
\end{figure}

Following these principles, we build \textit{MolPhenix}, a multi-modal \underline{mol}ecular \underline{phen}om\underline{ics} model addressing contrastive phenomolecular retrieval (Figure \ref{fig:main}). MolPhenix demonstrates large and consistent improvements in the presence of batch effects, generalizing across different concentrations, molecules, and activity thresholds.
Additionally, MolPhenix outperforms baseline methods in zero-shot setting, achieving 77.33\% top-1\% retrieval accuracies on active molecules, which corresponds to a \textbf{8.1$\times$} improvement over the previous state-of-the-art (SOTA) \cite{cloome}.
% This accounts for better data samples models and evaluation of metrics \db{This number will change when running the baselines with images rather than Ph-1}.

%% file: related_work.tex
\vspace{-0.25cm}
\section{Related Work}
\vspace{-0.25cm}

\textbf{Uni-modality Pretraining:} Self-supervised methods have demonstrated success across a variety of domains such as computer vision, natural language processing and molecular representations \cite{ssl_cookbook, gpt2, molecule_denoising_ssl}.
In vision, contrastive methods have been used to minimize distance in the model's latent space of two views of the same sample \cite{simclr2, ntxent, byol, moco}. 
Reconstruction objectives have also permeated computer vision, such as masked autoencoders (MAE).
MAEs typically utilize vision transformers to partition the image into learnable tokens and reconstruct masked patches \cite{mae, maespatio, maeunderstand, vit}. These methods have been extended to microscopy experimental data designed to capture cell morphology \cite{maestr, ph1}. Phenom1 utilizes a masked autoencoder with a ViT-L/8+ architecture and a custom Fourier domain reconstruction loss, yielding informative representations of phenomic experiments \cite{ph1, vitscaling}. From a representational perspective, Graph Neural Networks (GNN) have been used to predict molecular properties by reasoning over graph structures. A combination of reconstruction and supervised objectives have led to models generalizing to a diverse range of prediction tasks \cite{Mole, UniMol, molgps, grover}. Our work leverages uni-modal foundation models, which are used to generate embeddings of phenomic images and molecular graphs.

\textbf{Multi-Modal Objectives:} Multi-modal models combine samples from two or more domains, to learn rich representations and demonstrate flexible ways to predict sample properties \cite{clip, alayrac2022flamingo, kosmos1_vision_language}. Contrastive methods minimize distances between paired samples, traditionally in text-image domains. However, training these models is computationally expensive, requiring large datasets. Multiple contributions have allowed for a reduction in compute and data budgets by an order of magnitude. In \textit{LiT}, the authors demonstrate that utilizing uni-modal pretrained models for one or both modalities matches zero-shot performance with an order of magnitude fewer paired examples seen \cite{zhai2022lit}. Zhai et al. (2023) demonstrate that by replacing the softmax operation over cosine similarities with an element wise sigmoid loss, allows contrastive learners to improve performance under label noise regime \cite{siglip}. By using a uni-modal pre-trained modal to calculate similarities between samples from one of the modalities, Srinivasa et al. (2023) have demonstrated improved performance on zero-shot evaluation \cite{cwcl}.
% This work relaxes the equidistant negative pairs assumption, utilizing prior knowledge of sample similarities in one of the modalities \cite{cwcl}. 
In our work, we build along these directions in molecular phenomic multi-modal training.

\textbf{Molecular-Phenomic Contrastive Learning:} Prior works in contrastive phenomic retrieval have utilized the InfoNCE objective as a pre-training technique to construct uni-modal representations \cite{infonce}. 
% \textit{Nguyen} et al. (2023) propose a multi-modal objective trained on hand-engineered visual features and a GNN molecular encoder. The work demonstrates improved molecular property prediction over no pre-training \cite{morphology,cell_profiler}.
Recent methods have attempted to improve retrieval by using the InfoLOOB objective \cite{infoloob}. Specifically, CLOOME utilizes the InfoLOOB loss with hopfield networks for zero-shot retrieval on unseen data samples \cite{hopfield, cloome}. Our work is parallel to the above directions, demonstrating a significant increase in molecular-phenomic retrieval by building on algorithmic improvements from the multi-modality literature.

%% file: methodology.tex
% \begin{figure}[b]
%     \centering
%     \includegraphics[width=.95\textwidth]{figs/MolPhenix_v3.pdf}
%     \caption{Illustration of proposed guidelines when incorporated in our \textit{MolPhenix} contrastive phenomic retrieval framework. We address challenges by utilizing a uni-modal pretrained MAE \& MPNN models, inter-sample weighting with a dosage aware \texttt{S2L} loss, undersampling inactive molecules, and encoding molecular concentration.}
%     \label{fig:main}
% \end{figure}

\section{Methodology}

In this section, we explain key challenges facing phenomolecular retrieval and provide guidelines that are key methodological improvements behind the success of MolPhenix \ref{fig:main}.

\textbf{Preliminaries:} Our setting studies the problem of learning multi-modal representations of molecules and phenomic experiments of treated cells \cite{cloome}. The aim of this work is to learn a joint latent space which maps phenomic experiments of treated cells and the corresponding molecular perturbations into the same latent space. We consider a set of lab experiments $\mathcal{E}$ defined as the tuple ($\mathbf{X}, \mathbf{M}, \mathbf{C}, \mathbf{\Psi}$). Each experiment $\epsilon \in \mathcal{E}$ consists of data samples $\mathbf{x}_{i} \in \mathbf{X}$ (such as images) and perturbations $\mathbf{m}_{i} \in \mathbf{M}$ (such as molecules) which are obtained at a specific dosage concentration $\mathbf{c}_{i} \in \mathbf{C}$, while \(\mathbf{\psi} \in \mathbf{\Psi}\) denotes molecular activity threshold.

\begin{wrapfigure}{r}{0.5\textwidth}
\vspace{-.8cm}
\begin{center}
    \includegraphics[width=0.5\textwidth]{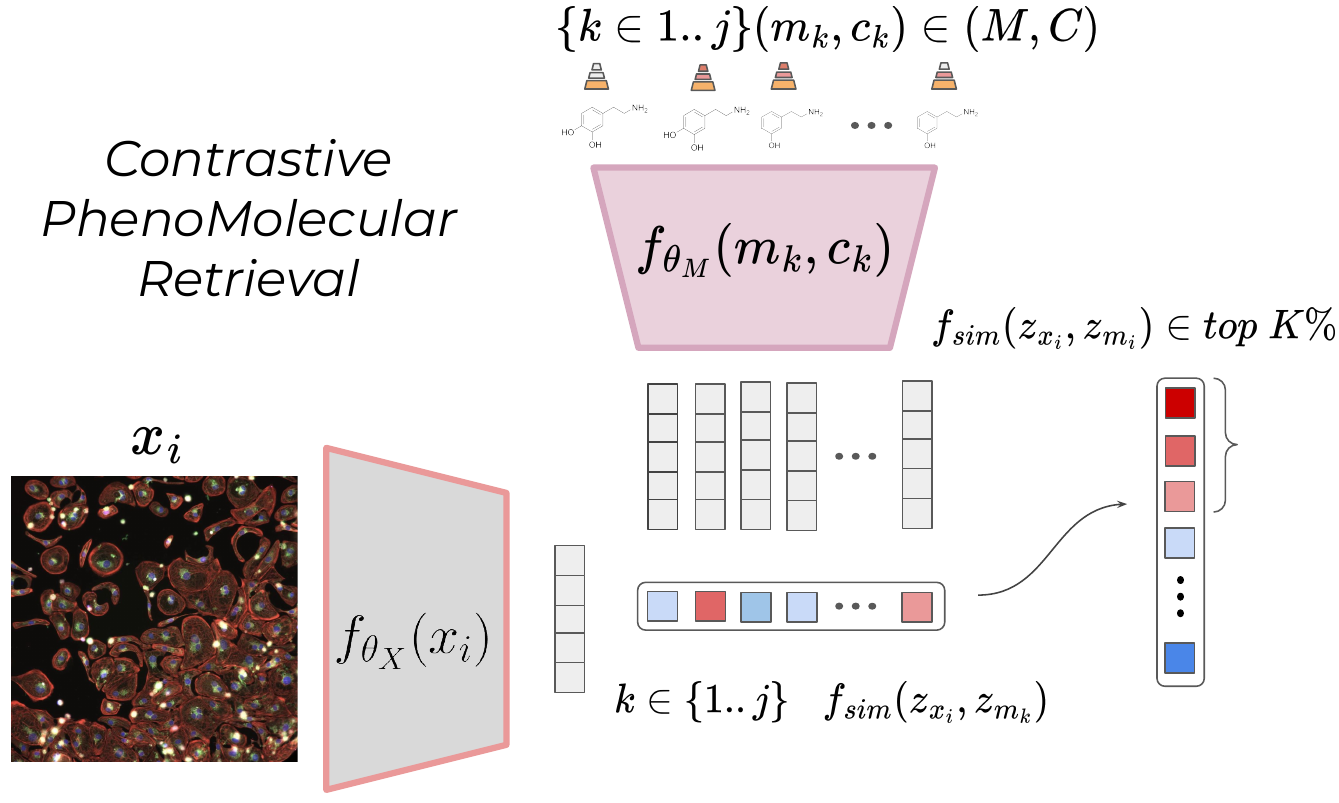}
    \caption{Illustration of the contrastive phenomolecular retrieval challenge. Image \(\mathbf{x}_i\) and a set of molecules and corresponding concentrations \((\mathbf{m}_k, \mathbf{c}_k)\) get mapped into a \(\mathbb{R}^d\) latent space. Their similarities get computed with \(f_{sim}\) and ranked to evaluate whether the paired perturbation appears in the top K\%.}
    \label{fig:contrastive_phenomolecular_retrieval}
\end{center}
\vspace{-.95cm}
\end{wrapfigure}

Figure \ref{fig:contrastive_phenomolecular_retrieval} describes the problem of contrastive phenomolecular retrieval, where for a single image \(\mathbf{x}_i\), the challenge consists of identifying the matching perturbation,  \(\mathbf{m}_i\), and concentration, \(\mathbf{c}_i\), used to induce morphological effects.
This can be accomplished in a zero-shot way by generating embeddings for \((\mathbf{m}_1, \mathbf{c}_1),... (\mathbf{m}_j, \mathbf{c}_j)\) and \(\mathbf{x}_i\) using functions \(f_{{\theta}_m}(\mathbf{m}, \mathbf{c})\), \(f_{{\theta}_x}(\mathbf{x})\) which map samples into \(\mathbb{R}^d\). 
Then, by defining a similarity metric between generated embeddings \(\mathbf{z}_{x_i}\) and \(\mathbf{z}_{m_i}\), \(f_{sim}\), we can rank \((\mathbf{m}_1, \mathbf{c}_1)... (\mathbf{m}_j, \mathbf{c}_j)\) based on computed similarities. An effective solution to the contrastive phenomolecular retrieval problem would learn \(f_{{\theta}_m}(\mathbf{m}, \mathbf{c})\) and \(f_{{\theta}_x}(\mathbf{x})\) that results in consistently high retrieval rates of \((\mathbf{m}_i, \mathbf{c}_i)\) used to perturb \(\mathbf{x}_i\).

% \begin{wrapfigure}{r}{0.5\textwidth}
% % \vspace{-.6cm}
% \begin{center}
%     \includegraphics[width=0.5\textwidth]{figs/contrastive_phenomolecular_v2.pdf}
%     \caption{Illustration of the contrastive phenomolecular retrieval challenge. Image \(\mathbf{x}_i\) and a set of molecules and corresponding concentrations \((\mathbf{m}_k, \mathbf{c}_k)\) get mapped into a \(\mathbb{R}^d\) latent space. Their similarities get computed with \(f_{sim}\) and ranked to evaluate whether the paired perturbation appears in the top K\%.}
%     \label{fig:contrastive_phenomolecular_retrieval}
% \end{center}
% \vspace{-0.75cm}
% \end{wrapfigure}

In practice, the image embeddings are generated using a phenomics microscopy foundation MAE model \cite{ph1, mae}. We use phenomic embeddings to marginalize batch effects, infer inter-sample similarities, and undersample inactive molecules. Activity is determined using consistency of replicate measurements for a given perturbation. For each sample, a $p$ value cutoff $\psi \in \mathbf{\Psi}$ is used to quantify molecular activity. Only molecules below the $p$ value cutoff $\psi$ are considered active.
% We explore a multitude of different methods to encode of molecular concentration (such as \texttt{one-hot} or \texttt{sigmoid}). 

Prior methods in multi-modal contrastive learning utilize the InfoNCE loss, and variants thereof \cite{infonce} to maximize the joint likelihood of \(\mathbf{x}_i\) and \(\mathbf{m}_i\). Given a set of $N \times N$ random samples \((\mathbf{x}_{1}, \mathbf{m}_{1}, \mathbf{c}_{1}), \cdots, (\mathbf{x}_{N}, \mathbf{m}_{N}, \mathbf{c}_{N})\) containing $N$ positive samples at $k^{\text{th}}$ index and $(N-1) \times N$ negative samples, optimizing Equation \ref{eq:infonce} maximizes the likelihood of positive pairs while minimizing the likelihood of negative pairs:

%  Figure out what the expectation is over
\begin{equation}
    \mathcal{L}_{\text{InfoNCE}} = - \frac{1}{N}\sum_{i=1}^N
    \left[ \log \frac{
    \exp(\langle \mathbf{z}_{x_i}, \mathbf{z}_{m_i} \rangle / \: \tau)
    }{
    \sum_{k = 1}^{N} \exp(\langle \mathbf{z}_{x_i}, \mathbf{z}_{m_k} \rangle \: / \tau) } + 
    \log \frac{
    \exp(\langle \mathbf{z}_{x_i}, \mathbf{z}_{m_i} \rangle / \: \tau)
    }{
    \sum_{k = 1}^{N} \exp(\langle \mathbf{z}_{m_i}, \mathbf{z}_{x_k} \rangle \: / \tau)
    } \right] 
    \label{eq:infonce}.
\end{equation}

Where \(\mathbf{z}_{x}\), \(\mathbf{z}_{m}\) correspond to phenomics and molecular embeddings respectively, \(\tau\) is softmax temperature,  and \(\langle \cdot \rangle\) corresponds to cosine similarity.

\subsection*{Challenge 1: Phenomic Pretraining and Generalization}

% Topic sentence
We find that using a phenomics foundation model to embed microscopy images allows for mitigation of batch effects, reduces the required number of paired data points, and improves generalization in the process. 
% Similar to text-image multi-modal training, contrastive phenomolecular retrieval involves mapping paired samples from distinct modalities to the same point in the learned latent space. 
 % Expand on data availability challenge
 While \texttt{CLIP}, a hallmark model in the field of text-image multi-modality, was trained on 400 million curated paired data points, there is an order of magnitude fewer paired molecular-phenomic molecule samples \cite{clip}. 
Cost and systematic pre-processing of data make large scale data generation efforts challenging, and resulting data is affected by experimental batch effects.
\textbf{Batch effects} induce noise in the latent space as a result of random perturbations in the experimental process, while biologically meaningful variation remains unchanged \cite{batcheffect2,batcheffect3}. 
% , making it difficult to find correlations within the data distribution \cite{batcheffect2,batcheffect3}. 
Limited dataset sizes and batch effects make it challenging for contrastive learners to capture molecular features affecting cell morphology, yielding low retrieval rates \cite{cloome}.

 % Wrap up with generalization challenges
%The above difficulties are further amplified in downstream generalization when evaluating model zero-shot generalization. Multi-modal molecular-phenomic models struggle to demonstrate positive transfer on unseen molecules, phenomic experiments and dosage concentrations. This occurs due to the inability of contrastive learners to capture inter-atomic relationships that drive the molecular effect on cells. 
% Instead, such models overfit to image modality priors and struggle to recuperate from locally suboptimal representations.

 %Our solution for taking a pre-trained model and multi stage generalization evaluation
We address data availability and generalization challenges by utilizing representations from a large \textbf{uni-modal pre-trained phenomic model}, $\theta_{\text{Ph}}$, trained to capture representations of cellular morphology. $\theta_{\text{Ph}}$ is pretrained on microscopy images using a Fourier modified MAE objective, utilizing the ViT-L/8 architecture with methodology similar to Kraus et al. (2024) \cite{mae,vit,ph1}. For simplicity in future sections, we refer to this model as \textit{Phenom1}. This pretrained model allows a drastic reduction in the required number of paired multi-modal samples \cite{zhai2022lit}. In addition, using phenomic representations alleviates the challenge of batch effects by averaging samples, \(\mathbf{z}_{x}\), generated with the same perturbation \(\mathbf{m}_i\) over multiple lab experiments \(\epsilon_i\). Averaging model representations \(\frac{1}{N} \Sigma^1_{i \in N} \mathbf{z}_{x_i}\) allows marginalizing batch effect induced by individual experiments.

%Guideline \ref{th:one} summarizes utilizing pretrained uni-modal models as a remedy.

\begin{tcolorbox}
\begin{theorem}
\label{th:one}
Utilizing pre-trained uni-modal encoder, $\theta_{\text{Ph}}$, can be used to reduce the number of paired data-points compared to training \(\theta\) without prior optimization. In addition, averaging phenomic embeddings \(\mathbf{z}_{x}\) from matched perturbations can alleviate batch effects.
\end{theorem}
\end{tcolorbox}

% While Phenom1 serves as an effective prior for phenomic experiments, certain models can be used to reason over molecular structures by leveraging additional inductive biases. Thus, we make use of features learned from GNNs trained on molecular property prediction \cite{gps++}. Specifically, we utilize a pretrained MPNN foundational model up to the order of 1B parameters for extracting molecular representations. We refer to this model as \textit{MolGPS} \cite{MolGPS}.

To reason over molecular structures, we make use of features learned from GNNs trained on molecular property prediction \cite{gps++}. We utilize a pretrained MPNN foundational model up to the order of 1B parameters for extracting molecular representations following a similar procedure to Sypetkowski et al. (2024) \cite{molgps}. We refer to this model as \textit{MolGPS}.

\subsection*{Challenge 2: Inactive Molecular Perturbations}

\begin{wrapfigure}{r}{0.45\textwidth}
\vspace{-1cm}
\begin{center}
    \includegraphics[width=0.42\textwidth]{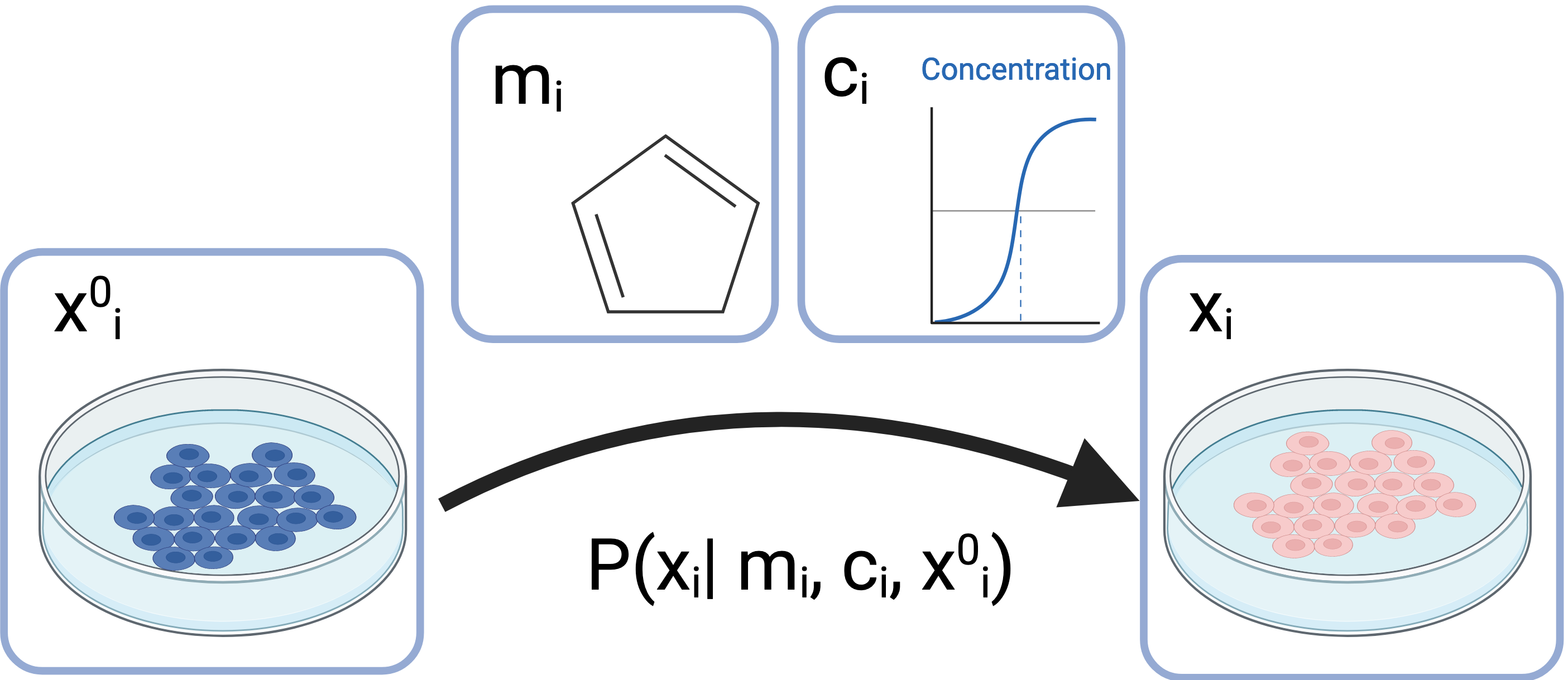}
    \caption{Data generation process of a phenomic experiment on cells \( \mathbf{x_i} \) with molecular perturbations \( \mathbf{m_i} \) and concentrations \( \mathbf{c_i} \).}
    \label{fig:no_molecular_effect}
\end{center}
\vspace{-0.6cm}
\end{wrapfigure}

% Describe the challenge intuitively and then provide a mathematical formulation
% We now focus on the challenge of perturbations arising from inactive molecules in cases where the molecule has no effect on the morphology of the perturbed cells. 
The phenomics-molecular data collection process can result in pairing of molecular structures with unperturbed cells in cases where the molecule has no effect on cell morphology (Figure \ref{fig:no_molecular_effect}) 

Since the morphological effects observed in cell \(\mathbf{x}_i\) is conditioned on the perturbation, in the absence of a molecular effect \(P(\mathbf{x_i} | \mathbf{x}^0_i, \mathbf{c_i}, \mathbf{m_i}) \sim P(\mathbf{x_i} | \mathbf{x}^0_i)\). In these samples, phenomic data will be independent, from paired molecular data, which results in misannotation under the assumption of data-pairs having an underlying biological relationship. We demonstrate how utilizing Phenom1 to undersample inactive molecules and learn continuous similarities between samples can alleviate this challenge.

% We use p-value thresholding 
To \textbf{undersample inactive molecules}, we extract the embeddings from Phenom1 and calculate the relative activity of each perturbation \((\mathbf{m}_i, \mathbf{c}_i) \in (\mathbf{M}, \mathbf{C})\). This is done using the rank of cosine similarities between technical replicates produced for a molecular perturbation against a null distribution. The null distribution is established by calculating cosine similarities from random pairs of Phenom1 embeddings generated with perturbation \((\mathbf{m}_j, \mathbf{c}_j), (\mathbf{m}_k, \mathbf{c}_k)\). Hence, we can compute a p-value and filter out samples likely to belong to the null distribution with an arbitrary threshold $\psi$.

% Use a specialized loss
In addition, by utilizing an inter-sample aware \textbf{\texttt{S2L} training objective}, the model can learn similarities between inactive molecules.
S2L is grounded in previous work which demonstrates improved robustness to label noise (SigLip) and learnable inter-sample associations (CWCL) \cite{siglip, cwcl}. 
Continuous Weighted Contrastive Loss (CWCL) provides better multi-modal alignment using a uni-modal pretrained model to suggest sample distances, relaxing the negative equidistant assumption present in InfoNCE \cite{cwcl}: 

\begin{equation}
    \mathcal{L}_{\text{CWCL}, \: \mathcal{M} \rightarrow \mathcal{X}} = - \frac{1}{N}\sum_{i=1}^N
    \left[ \frac{1}{\sum_{j=1}^{N} \mathbf{w}^{\mathcal{X}}_{i,j}}
    \sum_{j=1}^{N} \mathbf{w}^{\mathcal{X}}_{i,j} \log 
    \frac{ 
    \exp{\left(  \langle \mathbf{z}_{x_i},\mathbf{z}_{m_j} \rangle / \tau \right) }}
    { \sum_{k=1}^{N} \exp{ \left( \langle \mathbf{z}_{x_j}, \mathbf{z}_{m_k} \rangle / \tau \right) }} \right]. \label{eq:cwcl}
\end{equation}

CWCL weights logits with a continuous measure of similarity $\mathbf{w}^{\mathcal{X}}$, resulting in better alignment of embeddings $\mathbf{z}_{\mathbf{x}_{i}}$ and $\mathbf{z}_{\mathbf{m}_{j}}$ across modalities. In equation \ref{eq:cwcl}, \(\mathbf{w}^{\mathcal{X}}\) is computed using a within modality similarity function such as \( \mathbf{w}^{\mathcal{X}}_{i,j} = \langle z_{\mathbf{x}_i}, z_{\mathbf{x}_j} \rangle / 2 + 0.5 \). Note, the above formula is used only for mapping samples from modality \(\mathcal{M}\) to \(\mathcal{X}\) for which a pre-trained model \(\theta_{\text{Ph}}\)is available.

Another work, SigLIP, demonstrates robustness to label noise and reduces computational requirements during contrastive training \cite{siglip}. It does so by avoiding computation of a softmax over the entire set of in-batch samples, instead relying on element-wise sigmoid operation:

\begin{equation}
    \mathcal{L}_{\text{SigLIP}} = 
    - \frac{1}{N}
    \sum_{i=1}^N 
    \sum_{j=1}^N
    \left[ \log 
    \frac{1}{1 + \exp{ \left( \mathbf{y}_{i,j}(-\alpha \: \langle \mathbf{z}_{\mathbf{x}_{i}}, \mathbf{z}_{\mathbf{m}_{j}} \rangle + b) \right) }}
    \right]. \label{eq:siglip}
\end{equation}

In equation \ref{eq:siglip}, \(\alpha\) and \(b\) are learned, calibrating the model confidence conditioned on the ratio of positive to negative pairs. \(\textbf{y}_{i,j}\) is set to 1 if \(i = j\) and -1 otherwise. 

Inspired by prior works, we introduce S2L for molecular representation learning, which leverages inter-sample similarities and robustness to label noise to mitigate weak or inactive perturbations. 

\begin{equation}
    \mathcal{L}_{\text{S2L}} =  
    - \frac{1}{N}
    \sum_{i=1}^N 
    \sum_{j=1}^N
    \log \left[ 
    \frac{\mathbf{w}^{\mathcal{X}}_{i,j}}
    {1 + \exp{ \left( -\alpha \mathbf{z}_{\mathbf{x}_{i}}.\mathbf{z}_{\mathbf{m}_{j}} + b) \right) }} +
    \frac{(1 - \mathbf{w}^{\mathcal{X}}_{i,j})}
    {1 + \exp{ \left( \alpha \mathbf{z}_{\mathbf{x}_{i}}.\mathbf{z}_{\mathbf{m}_{j}} + b) \right) }} 
    \right]. \label{eq:s2l}
\end{equation}

In the equation above, \(\mathbf{z}_{\mathbf{x}_i}\) and \(\mathbf{z}_{\mathbf{m}_j}\) correspond to latent representations of images and molecules, respectively. \(\alpha\) and \(\
b\) correspond to learnable temperature and bias parameters for the calibrated sigmoid function. \(\mathbf{w}^{\mathcal{X}}_{ij}\) is an inter-sample similarity function computed from images using the pretrained model \(\theta_{\text{Ph}}\). To compute \(\mathbf{w}^{\mathcal{X}}_{i,j}\), we use the arctangent of L2 distance instead of cosine similarity, as was the case for Equation \ref{eq:cwcl} (more details in Appendix \ref{sec:s2l_distance_appendix}). Intuitively, S2L can be thought of as shifting from a multi-class classification to a soft multi-label problem. In our problem setting, the labels are continuous and determined by sample similarity in the phenomics space.

\begin{tcolorbox}
\begin{theorem}
\label{th:two}
When training a molecular-phenomic model, mitigating the effect of inactive molecules in training data distribution can be carried out by undersampling inactive molecules and using an inter-sample similarity aware, \texttt{S2L} loss (equation \ref{eq:s2l}).
\end{theorem}
\end{tcolorbox}

\subsection*{Challenge 3: Variable Concentrations}

% Topic sentence indicating that molecular impact is a function of molecule dosage 
% As well as perturbation identity
Perturbation effect on a cell is determined by both molecular structure and corresponding concentration \cite{carcinogenicity_dosage}. A model capturing molecular impact on cell morphology must be able to generalize across different doses, since variable concentrations can correspond to different data distributions.

We note that providing concentrations $\mathbf{c}_{i}$ as input to the model would benefit performance, as this would indicate the magnitude of molecular impact. However, we find that simply concatenating concentrations does not result in effective training due to its compressed dynamic range. To that end, we add concentration information in two separate ways: \textit{implicit} and \textit{explicit} formulations.

We add \textbf{implicit concentration}  as molecular perturbation classes by using the S2L loss (Equation \ref{eq:s2l}) to treat perturbation \(\mathbf{m}_i\) with concentrations \(\mathbf{c}_i\) and \(\mathbf{c}_j\) as distinct classes. This pushes samples apart in the latent space proportionally to similarities between phenomic experiments.

We add \textbf{explicit concentration}  $c_{i}$ by passing it to the molecular encoder. We explore different formulation for dosage concentrations, $\mathbf{f'}(c_{i})$, where $\mathbf{f'} \text{ maps } \mathbf{c_i} \rightarrow \mathbb{R}$. 
Encoded representations $\mathbf{f'}(c_{i})$ are concatenated at the initial layer of the model.
% While one may choose to learn the mapping $\mathbf{f'}$, we found simpler functional encodings (such as \texttt{one-hot} and \texttt{logarithm}) to work well in practice. % Guideline \ref{th:three} presents our remedy for tackling variable concentration training.
 We find simple functional encodings $\mathbf{f'}$ (such as \texttt{one-hot} and \texttt{logarithm}) to work well in practice. 

\begin{tcolorbox}
\begin{theorem}
\label{th:three}
When training a molecular-phenomic model, conditioning on an (implicit and explicit) representation of concentration $\mathbf{f'}(\mathbf{c}_{i})$ aids in capturing molecular impacts on cell morphology and improves generalization to previously unseen molecules and concentrations.
\end{theorem}
\end{tcolorbox}

%% file: experiments.tex
\section{Experimental Setup}
% Add a roadmap to the first paragraph describing what we actually do. 
In this section, we describe evaluation datasets used, and descriptions of the underlying data modalities. 
To assess phenomolecular retrieval, we use 1\% recall metric unless stated otherwise, as it allows direct comparison between datasets with different number of samples. Additional implementation and evaluation details can be found in Appendix \ref{sc:appa}.

% \subsection{Addressing Challenges in Contrastive Phenomic Retrieval}

\paragraph{Datasets:} Our training dataset consists of fluorescent microscopy images paired with molecular structures and concentrations, which are used as perturbants. 
We assess models' phenomolecular retrieval capabilities on three datasets of escalating generalization complexity. 
First dataset, consisting of unseen microscopy images and molecules present in the training dataset. 
Second, a dataset consisting of previously unseen phenomics experiments and molecules split by the corresponding molecular scaffold.
Finally, we evaluate on an open source dataset with a different data generating distribution \cite{rxrx3}. 
% Make a note that this is 0-shot prediction
In the case of the latter two datasets, the model is required to perform zero-shot classification, as it has no access to those molecules in the training data.
This requires the model to reason over molecular graphs to identify structures inducing corresponding cellular morphology changes.
Using methodology described in guideline \ref{th:two} we report retrieval results for all molecules as well as on an active subset.
Finally, all datasets are comprised of molecular structures at multiple concentrations ($.01, .1, 1.0, 10$, etc.) Additional details regarding the datasets can be found in Appendix \ref{sc:dataset}.

\paragraph{Modality Representations:} In our evaluations, we consider different representations for molecular perturbations and phenomic experiments and quantitatively evaluate their impact.

% Use minipage to control layout
\begin{itemize}[leftmargin=*]
  \item \texttt{Images}: Image encoders utilize 6-channel fluorescent microscopy images of cells representing phenomic experiments. Images are 2048 $\times$ 2048 pixels, capturing cellular morphology changes post molecular perturbation. We downscale each image to 256 $\times$ 256 using block mean downsampling.
  \item \texttt{Phenom1}: We characterize phenomic experiments by embedding high resolution microscopy images in the latent space of a phenomics model \(\theta_{Ph}\) as described in guideline \ref{th:one}.
  \item \texttt{Fingerprints}: Molecular fingerprints utilize RDKIT \cite{landrum2013rdkit}, MACCS \cite{maacs} and MORGAN3 \cite{morgan} bit coding, which represent binary presence of molecular substructures. Additional information such as atomic identity, atomic radius and torsional angles are included in the fingerprint representations.
  \item \texttt{MolGPS}: We generate molecular representations from a large pretrained GNN. Specifically, we obtain molecular embeddings from a 1B parameter MPNN \cite{gps++}.
\end{itemize}

\begin{table}[b]
    \centering
    \caption{Impact of pre-trained Phenom1 and MolGPS on CLOOME and MolPhenix for a matched number of seen samples (Top), where we observe an \textbf{8.1} $\times$ improvement of MolPhenix over the CLOOME baseline for active unseen molecules. SOTA results trained with a higher number of steps by utilizing the best hyperparameters (Bottom *). We note that MolPhenix's  main components such as S2L and embedding averaging relies on having a pre-trained uni-modal phenomics model.}
    \vspace{0.25cm}
    \label{tab:ScratchVsPhenom1}
    \begin{adjustbox}{max width=\textwidth}
    \begin{tabular}{llcccccc}
    \toprule
    \multicolumn{2}{c}{} & \multicolumn{3}{c}{\textbf{Active Molecules}} & \multicolumn{3}{c}{\textbf{All Molecules}} \\
     \cmidrule(lr){3-5} \cmidrule(lr){6-8}
    {Method} & {Modality} & \thead{Unseen \\ Im.} & \thead{{Unseen} \\ {Im. + Mol.}} & \thead{Unseen \\ Dataset} & \thead{Unseen \\ Im.} &  \thead{{Unseen} \\ {Im. + Mol.}} & \thead{Unseen \\ Dataset}\\
    \toprule
    CLOOME  & Images \& Multi-FPS& $.0756 \pm .0042$ & $.0787 \pm .0065$ & $.0528 \pm .0057$ & $.0547 \pm .0028$ & $.0661 \pm .0020$ & $.0223 \pm .0014$ \\
    CLOOME & Phenom1 \& Multi-FPS& $.4659 \pm .0042$ & $.5057 \pm .0014$ & $.2065 \pm .0146$ & $.3009 \pm .0053$ & $.2474 \pm .0013$ & $.1337 \pm .0045$ \\
    MolPhenix & Phenom1 \& Multi-FPS& $\mathbf{.7807 \pm .0025}$ & $.6365 \pm .0014$ & $.3545 \pm .0097$ & $\mathbf{.5253 \pm .0029}$ & $\mathbf{.3655 \pm .0017}$ & $.2163 \pm .0021$\\
    MolPhenix & Phenom1 \& MolGPS & $.7646 \pm .0014$ & $\mathbf{.6387 \pm .0056}$ & $\mathbf{.4160 \pm .0016}$ & $.5012 \pm .0002$ & $.3511 \pm .0004$ & $\mathbf{.2508 \pm .0026}$ \\
    \toprule
    MolPhenix* & Phenom1 \& MolGPS  & $\mathbf{.9689 \pm .0017}$ & $\mathbf{.7733 \pm .0036}$ & $\mathbf{.5860 \pm .0082}$ & $\mathbf{.5583 \pm .0007}$ & $\mathbf{.3824 \pm .0016}$ & $\mathbf{.2809 \pm .0060}$ \\
    \end{tabular}
    \end{adjustbox}
    % \vspace{.25cm}
\end{table}

\begin{table}[t]
    \centering
    \caption{Top-1\% recall accuracy with use of the proposed MolPhenix guidelines, such as Phenom1 and embedding averaging.
    % S2L demonstrates higher retrieval rates through use of inter-sample similarities and implicit dosage. 
    We omit explicit concentration from this experiment.}
    \vspace{.2cm}
    \label{tab:loss}
    \resizebox{1\textwidth}{!}{
    \begin{tabular}{lcccccc}
    \toprule
    \multicolumn{1}{c}{} & \multicolumn{3}{c}{\textbf{Active Molecules}} & \multicolumn{3}{c}{\textbf{All Molecules}} \\
    \cmidrule(lr){2-4} \cmidrule(lr){5-7}
    {Loss} & \thead{Unseen \\ Images} &  \thead{{Unseen} \\ {Im. + Mol.}} & \thead{Unseen  \\ Dataset} & \thead{Unseen \\ Images} &  \thead{{Unseen} \\ {Im. + Mol.}} & \thead{Unseen \\ Dataset}\\
    \midrule
    CLIP & $.3373 \pm .0043$ & $.4228 \pm .0008$ & $.1514 \pm .0038$ & $.1761 \pm .0043$ & $.1867 \pm .0022$ & $.0734 \pm .0022$\\
    Hopfield-CLIP & $.2578 \pm .0042$ & $.3559 \pm .0042$ & $.1256 \pm .0092$ & $.1531 \pm .0046$ & $.1709 \pm .0029$ & $.0673 \pm .0020$ \\
    InfoLOOB & $.3351 \pm .0011$ & $.4206 \pm .0031$ & $.1563 \pm .0028$ & $.1746 \pm .0003$ & $.1860 \pm .0029$ & $.0745 \pm .0019$ \\
    CLOOME & $.3572 \pm .0026$ & $.4348 \pm .0039$ & $.1658 \pm .0063$ & $.1968 \pm .0029$ & $.2005 \pm .0026$ & $.0911 \pm .0022$\\
    DCL & $.6363 \pm .0025$ & $.6177 \pm .0047$ & $.3184 \pm .0087$ & $.3277 \pm .0047$ & $.2562 \pm .0008$ & $.1364 \pm .0067$ \\
    CWCL & $.7091 \pm .0045$ & $.6529 \pm .0020$ & $.3556 \pm .0094$ & $.3635 \pm .0064$ & $.2696 \pm .0019$ & $.1526 \pm .0058$ \\
    SigLip  & $.7763 \pm .0045$ & $.6401 \pm .0065$ & $.3396 \pm .0042$ & $.3729 \pm .0039$ & $.2544 \pm .0014$ & $.1470 \pm .0038$ \\
    S2L (ours) &  $\mathbf{.9097 \pm .0020}$ & $\mathbf{.6759 \pm .0012}$ &  $\mathbf{.4181 \pm .0012}$ & $\mathbf{.4688 \pm .0009}$ & $\mathbf{.2852 \pm .0001}$ & $\mathbf{.1838 \pm .0007}$\\
    \bottomrule
    \end{tabular}
    % \vspace{-.7cm}
}
\end{table}
\begin{table}[t]
    \vspace{-.5cm}
    \centering
    \caption{Top-1\% recall accuracy across different concentration encoding choices with use of the proposed MolPhenix guidelines, such as Phenom1 and embedding averaging. 
    % Explicitly providing of one-hot encoding of concentrations leads to higher retrieval rates.
    }
    \vspace{.2cm}
    \resizebox{1\textwidth}{!}{
    \begin{tabular}{cccccccc}
    \toprule
    \multicolumn{2}{c}{} & \multicolumn{3}{c}{\textbf{Active Molecules}} & \multicolumn{3}{c}{\textbf{All Molecules}} \\
    \cmidrule(lr){3-5} \cmidrule(lr){6-8}
    \thead{Implicit  \\ Concentration} & \thead{Explicit \\ Concentration} & \thead{Unseen \\ Im.} &  \thead{{Unseen} \\ {Im. + Mol.}} & \thead{Unseen \\ Dataset} & \thead{Unseen \\ Im.} &  \thead{{Unseen} \\ {Im. + Mol.}} & \thead{Unseen \\ Dataset} \\
    \midrule
    \xmark & \xmark  & $.7350 \pm .0071$ & $.6509 \pm .0104$ & $.3333 \pm .0004$ & .$3610 \pm .0025$ & $.2668 \pm .0034$ & $.1532 \pm .0007$ \\
    \cmark & \xmark & $.9097 \pm .0020$ & $.6759 \pm .0012$ &  $.4181 \pm .0012$ & $.4688 \pm .0009$ & $.2852 \pm .0001$ & $.1838 \pm .0007$ \\
    \cmark & sigmoid  & $.9423 \pm .0011$ & $.7155 \pm .0016$ &  $.4573 \pm .0022$ & $.5071 \pm .0024$ & $.3441 \pm .0026$ & $.2144 \pm .0026$\\
    \cmark & logarithm  & $.9426 \pm .0066$ & $.7451 \pm .0050$ & $.4727 \pm .0056$ & $.5183 \pm .0027$ & $.3700 \pm .0036$ & $.2275 \pm .0032$  \\
    \cmark & one-hot  & $\mathbf{.9430 \pm .0029}$ & $\mathbf{.7490 \pm .0052}$ & $\mathbf{.4850 \pm .0020}$ & $\mathbf{.5433 \pm .0030}$ & $\mathbf{.3819 \pm .0032}$ & $\mathbf{.2384 \pm .0049}$ \\
    \bottomrule
    \end{tabular}
    }
    % \vspace{.25cm}
    \label{tab:summary_cross_dosage}
    \vspace{-.25cm}
    % \vspace{-.75cm}
\end{table}

% \begin{wrapfigure}{r}{.3\textwidth}
% \vspace{-.9cm}
% \begin{center}
%     \includegraphics[width=.3\textwidth]{figs/ScratchVsPretrained.pdf}
%     \caption{sffv}
%     \label{fig:contrastive_phenomic_retrieval}
% \end{center}
% \vspace{-.25cm}
% \end{wrapfigure}

% \begin{table}[t]
%     \centering
%     \begin{adjustbox}{max width=.5\textwidth}
%     \begin{tabular}{cccccccc}
%          Method &  Modality & Unseen Images & Unseen Images + & Unseen Dataset\\
%          &   &   & Unseen Molecules &  \\
%         \toprule
%         CLOOME  & Images & - & - & - \\
%         CLOOME & Phenom1 & - & - & - \\
%         MolPhenix & Phenom1 & .9430 & .7490 &  .4850 \\
%         MolPhenix & Phenom1 & \textbf{.9430} & \textbf{.7514} & \textbf{.5577} \\
%          & \& MolGPS &  &  &  \\
%     \end{tabular}
%     \end{adjustbox}
%     \caption{Caption}
%     \label{tab:my_label}
% \end{table}

% \begin{wrapfigure}{r}{0.45\textwidth}
%     \centering
%     \vspace{-1.2cm}
%     \includegraphics[width=\linewidth]{figs/ScratchVsPretrained.pdf}
%     \caption{Comparison of training phenomic encoder from scratch and utilizing pre-trained Phenom1 unseen dataset. X-axis plotted on logarithmic scale.}
%     \vspace{-1.1cm}
%     \label{fig:scratchVsPhenom1}
    
% \end{wrapfigure}

\section{Results and Discussion}

To evaluate the effectiveness of Guidelines \ref{th:one}, \ref{th:two}, and \ref{th:three} we carry out evaluation in two different settings: (1) cumulative concentrations, and (2) held-out concentrations, testing the models' ability to generalize to new molecular doses. Finally, we perform comprehensive ablations testing model performance with varying data, model, and optimization parameters. 
The comprehensive set of results can be found in Tables \ref{tab:appendix_cross_dose_results}, \ref{tab:appendix_within_dose},
\ref{tab:appendix_cross_dose_results_all_mols}, and \ref{tab:appendix_whitin_dose_results_all_mols}.

\subsection{Evaluation on cumulative concentrations:}

\begin{wrapfigure}{r}{0.45\textwidth}
    \centering
    % \vspace{-1.2cm}
    \includegraphics[width=\linewidth]{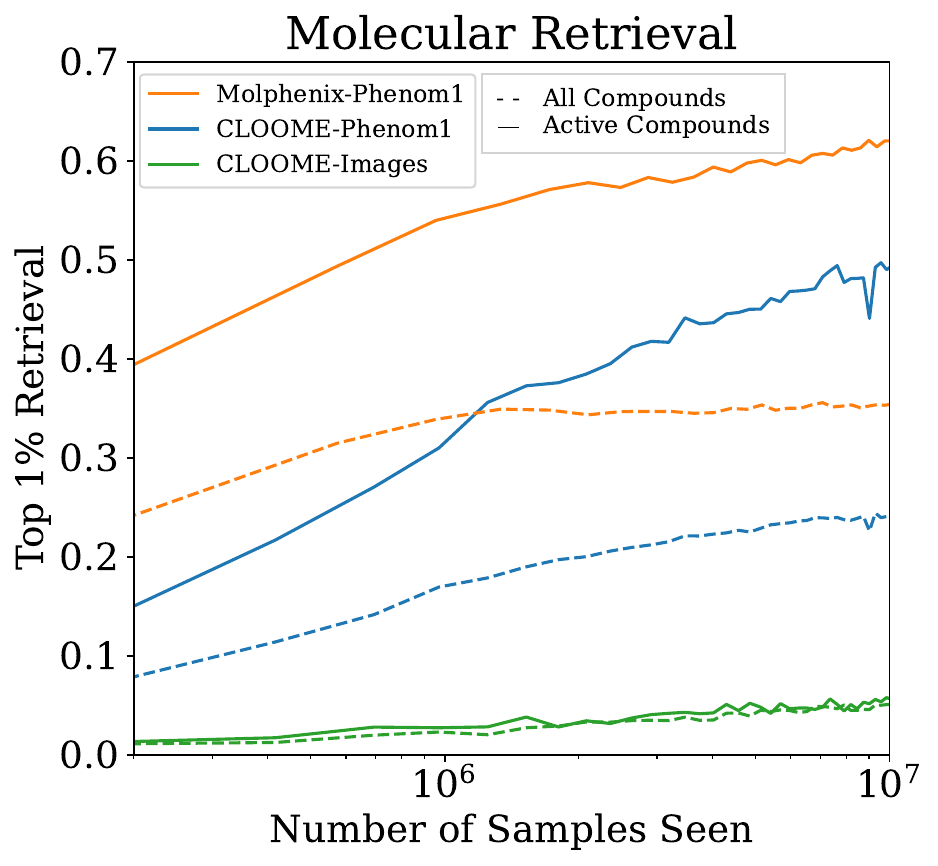}
    \caption{Comparison of training phenomic encoder from scratch and utilizing pre-trained Phenom1 unseen dataset. X-axis plotted on logarithmic scale.}
    % \vspace{-1.1cm}
    \label{fig:scratchVsPhenom1}
    
\end{wrapfigure}

We demonstrate improvements in phenomolecular recall due to usage of a phenomics pre-trained foundation model, identify that MolPhenix set of design choices results in higher final performance, and more data efficient learning. 
Figure \ref{fig:scratchVsPhenom1} demonstrates recall accuracy on all molecules and an active subset for CLOOME and MolPhenix models, as a function of training samples seen.

We observe a large performance gap between models trained on Phenom1 embeddings as opposed to images, emphasizing the utility of using a pre-trained encoder for microscopy images (Table \ref{tab:ScratchVsPhenom1}).
We note that provision of Phenom1 (CLOOME-Phenom1 Vs CLOOME-Images) significantly improves both active and all molecule retrieval by $ \mathbf{5.69\times}$ and, $\mathbf{4.75 \times}$ respectively (Table \ref{tab:ScratchVsPhenom1}). 

Furthermore, we identify that while all molecules retrieval stagnates throughout training, the performance on an active subset keeps improving, underscoring the importance of identification of the active subset.
% Mention why we thinkg this is important - active molecules are the most important ones 
% This validates our choice for reporting performance on active molecules, demonstrating continuous improvement on the active subset throughout training.
% Mention there is a huge gap between CLOOME-Phenom1 and MolPhenix
Finally, we compare CLOOME and MolPhenix trained using Phenom1 embeddings and find there is a consistent retrieval performance gap, throughout training, with a \textbf{1.26} \(\times\) final improvement (Figure \ref{fig:scratchVsPhenom1}, Table \ref{tab:ScratchVsPhenom1}). 
Compared to CLOOME~\cite{cloome} trained directly on images, MolPhenix achieves an average improvement of $\mathbf{8.78 \times}$ on active molecules on the unseen dataset. 
These results verify the effectiveness of Guideline \ref{th:one} in accelerating training, and the importance of Guidelines \ref{th:two} and \ref{th:three} in recall improvements over CLOOME. 
% \db{Note that MolPhenix is also $\mathbf{8.4 \times}$ faster than CLOOME as detailed in appendix \ref{sc:compute}.}
% We further investigate the impact of individual design choices by methodically comparing recall performances.

% Compared to previous SOTA models trained from scratch (such as CLOOME [38])
% (such as CLOOME~\cite{cloome})
We evaluate the impact of different loss objectives on the proposed MolPhenix training framework. 
Table \ref{tab:loss} presents top-1\% retrieval accuracy across different contrastive losses utilized to train molecular-phenomics encoders on cumulative concentrations. 
Compared to prior methods, the proposed \texttt{S2L} loss demonstrates improved retrieval rates in cumulative concentration setting.
Label noise and inter-sample similarity aware losses such as CWCL and SigLip also demonstrate improved performance.
% These findings underscore the efficacy of Guideline \ref{th:two} in mitigating noise originating from inactive molecular perturbations. 
The effectiveness of S2L can be attributed to smoothed inter-sample similarities and implicit concentration information.

Finally, in Table \ref{tab:summary_cross_dosage}, we observe recall improvements when considering both molecular structures and concentration.
%To disentangle effects of implicit and explicit concentration encoding, we systematically assess contributions of each. 
We note the importance of the addition of implicit concentration, further confirming the importance of considering molecular effects at different concentrations as different classes.
Explicitly encoding molecular concentration with \texttt{one-hot}, \texttt{logarithm} and \texttt{sigmoid} yields improved recall performance, where \texttt{one-hot} performs the best in a cumulative concentration setting. These findings verify the efficacy of implicit and explicit concentration encoding outlined in Guideline \ref{th:three}.

\begin{table}[H]
\vspace{-.5cm}
    \centering
    \begin{minipage}[t]{.4\textwidth}
        \centering
        \caption{Top-1\% recall accuracy of different loss objectives while using the proposed MolPhenix guidelines, such as Phenom1 and embedding averaging. 
        % S2L demonstrates higher retrieval rates by tackling inactive molecular perturbations.
        }
        \vspace{.1cm}
        \label{tab:heldout_loss}
        \begin{adjustbox}{max width=\textwidth}
            \begin{tabular}{lcccccc}
                \toprule
                {Loss} & \thead{Unseen \\ Im.} &  \thead{{Unseen} \\ {Im. + Mol.}} & \thead{Unseen \\ Dataset}  \\
                \toprule
                CLIP &  .2109 & .2425 & .1519 \\
                Hopfield-CLIP & .1581 & .2034 & .1198 \\
                InfoLOOB & .2122 & .2496 & .1501 \\
                CLOOME & .2164 & .2461 & .1479 \\
                DCL & .4717 & .4027 & .2841 \\
                CWCL  & .5731 & .4403 & .3232 \\
                SigLip & .5718 & .4217 & .3021 \\
                S2L (ours) & \textbf{.8334} & \textbf{.4615} & \textbf{.3792} \\
                \bottomrule
            \end{tabular}
        \end{adjustbox}
    \end{minipage}\hfill
    \begin{minipage}[t]{.58\textwidth}
        \centering
        \caption{Top-1\% recall accuracy across different concentration encoding choices while using the proposed MolPhenix guidelines, such as Phenom1 and embedding averaging.}
        \vspace{.5cm}
        \label{tab:heldout_conditioning}
        \begin{adjustbox}{max width=\textwidth}
            \begin{tabular}{cccccc}
                \toprule
                \thead{Implicit  \\ Concentration} & \thead{Explicit \\ Concentration} & \thead{Unseen \\ Im.} &  \thead{{Unseen} \\ {Im. + Mol.}} & \thead{Unseen \\ Dataset} \\
                \toprule
                \xmark & \xmark & .5942 & .4315 & .3129 \\
                \cmark & \xmark & \textbf{.8334} & .4615 & \textbf{.3792} \\
                \cmark & sigmoid & .8256 & \textbf{.4692} & .3765 \\
                \cmark & logarithm & .7953 & .4466 & .3664 \\
                \cmark & one-hot & .7489 & .4088 & .3379 \\
                \bottomrule
            \end{tabular}
        \end{adjustbox}
    \end{minipage}
    \vspace{.1cm}
    \caption*{Results are averaged across experiments for each dropped concentration, and across three seeds. Recall is reported for active molecules, while the results for all molecules can be found in Table \ref{tab:appendix_whitin_dose_results_all_mols}.}
    % \vspace{-.7cm}
\end{table}

\subsection{Evaluation on held-out concentrations:}

Next, we evaluate recall on held-out concentrations to obtain a measure of generalization performance. 
This evaluation allows us to capture the utility of our models for prediction of unseen concentrations, hence resembling \textit{in-silico} testing. 
We omit concentrations from the training set and evaluate recall at the excluded data, where we observe a drop in retrieval performance for unseen concentrations. 
Similar to cumulative concentration results, we find that using \texttt{S2L} improves recall over other losses and outperforms CLOOME by up to 126\% (Table \ref{tab:heldout_loss}).
While \texttt{one-hot} encoding exhibits significant improvements in cumulative concentrations, its expressivity on unseen concentrations is limited (Table \ref{tab:heldout_conditioning}) and \texttt{sigmoid} encoding provides a sufficient representation of concentration. 

\subsection{Ablation Studies}

\begin{figure}[h]
    \centering
    % \vspace{-1cm}
    \includegraphics[width=.85\textwidth]{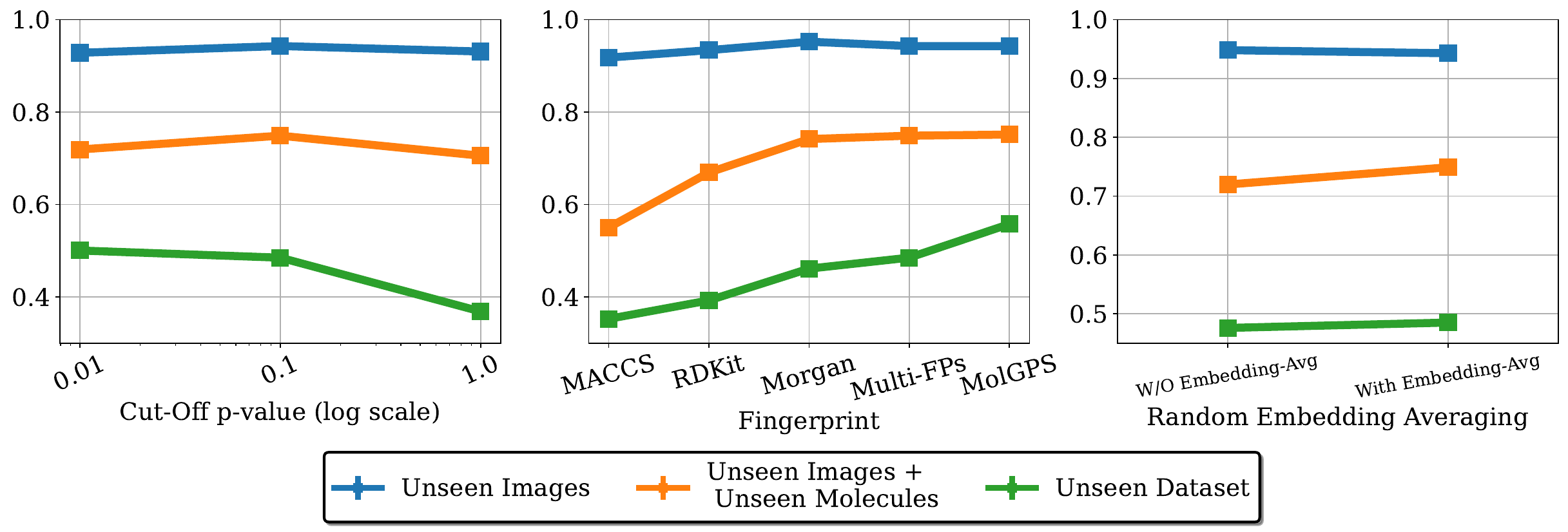}
    \caption{Ablations of top-1 \% recall accuracy with \textbf{(bottom-left)} cutoff $p$ value, \textbf{(bottom-center)} fingerprint type, and \textbf{(bottom-right)} embedding averaging. 
    % Compact embedding sizes from pretrained models, larger number of parameters, larger batch sizes, lower cutoff p-values, pretrained MolGPS fingerprints and presence of random batch averagin improving retrieval of our MolPhenix framework.
    }
    \vspace{-.4cm}
    \label{fig:ablations_main}
\end{figure}

We assess the importance of our design decisions by conducting an ablation study over our proposed guidelines.
Figure \ref{fig:ablations_main} presents the variation of top-1\% recall accuracy across key components such as cutoff $p$ value, fingerprint type, and embedding averaging.
We observe that employing a lower cutoff $p$ value yields improved generalization for unseen dataset, while employing a higher cutoff appears to be optimal for unseen images + unseen molecules. 
% In addition, we examine the influence of fingerprint types on retrieval performance. 
% Our ablations confirm that combining diverse molecular fingerprints from pretrained models outperforms their individual usage. 
For molecular structure representations, we find that using embeddings from the large pretrained MPNN graph based model (e.g., MolGPS) surpasses traditional fingerprints. 
Finally, utilization of embedding averaging demonstrates improved recall. 
More ablations over model size, projector dimension, and batch size can be found in 
Appendix \ref{sc:ablation}.

%  \begin{table}[H]
%     \centering
%   \begin{adjustbox}{max width=\textwidth}
%     \begin{tabular}{c|c|c|c|c|c|c}
%         Method & Dosage Information & Explicit dosage & Modality & Unseen Images & Unseen Images + Unseen Molecules & Unseen Dataset (0-shot) \\
%         \toprule
%         MolPhenix (ours) & \cmark & one-hot & Phenom1 \& MolGPS & .9689 & .7733 & .5860\\
%         \bottomrule
%     \end{tabular}
%     \end{adjustbox}
%     \vspace{.5cm}
%     \caption{State-of-the-art results results. \db{Add results from the CLOOME and InfoLOOB with images and only explicit dosage.}}
%     \label{tab:sota}
% \end{table}

%% file: conclusion.tex
\section{Conclusion}

In this work, we investigate the problem of \textit{contrastive phenomolecular retrieval} by constructing a joint multi-modal embedding of phenomic experiments and molecular structures. 
We identify a set of challenges afflicting molecular-phenomic training and proposed a set of guidelines for improving retrieval and generalization. 
Empirically, we observed that contrastive learners demonstrate higher retrieval rates when using representations from a high-capacity uni-modal pretrained model. 
Use of inter-sample similarities with a label noise resistant loss such as \texttt{S2L} allows us to tackle the challenge of inactive molecules. 
% Downsampling inactive molecules further improved performance. 
Finally, adding implicit and explicit concentrations allows models to generalize to previously unseen concentrations. 
MolPhenix demonstrates an \textbf{8.1$\times$} improvement in zero shot retrieval of active molecules over the previous state-of-the-art, reaching 77.33\% in top-1\% accuracy. In addition, we conduct a preliminary investigation on MolPhenix's ability to uncover biologically meaningful properties (activity prediction, zero-shot biological perturbation matching, and molecular property prediction in Appendix \ref{sec:molecular_activity}, \ref{sec:zero_shot_validaiton}, and \ref{sec:tdc_polaris}, respectively.). We expect a wide range of applications for MolPhenix, particularly in drug discovery. While there's a remote chance of misuse for developing chemical weapons, such harm is unlikely, with our primary focus remaining on healthcare improvement. 

\textbf{Limitations and Future Work:} While our study covers challenges in phenomolecular recall, we leave three research directions for future work. 
(1) Future investigations could consider studying additional modalities such as text, genetic perturbations and chemical multi-compound interventions. 
(2) While we propose and evaluate our guidelines on previously conducted phenomic experiments, we note that a rigorous evaluation would evaluate model predictions in a wet-lab setting. 
(3) In addition, our work makes the assumption that the initial unperturbed cell state \(x^0_i\) can be marginalized by utilizing a single cell line with an unperturbed genetic background. Future works can relax this assumption, aiming to capture innate intercellular variation.

% \begin{ack}

% We thank the broader Valence Labs team and Recursion Pharmaceuticals for support on the project. We thank Berton Earnshaw, Jason Hartford and Emmanuel Noutahi for providing valuable feedback. 

% \end{ack}

%% file: appendix.tex
\appendix

\newpage

\section{Assumption of the Initial Cell State}
There is an important distinction between phenomics - molecule and text - image contrastive training although there are initial similarities. In the text - image domain the two modalities are directly generated by the same latent variable which is the underlying semantic class. Whereas in phenomics - molecule, the observed phenomics variable is actually conditioned on molecular structure and the initial state. There are two important conclusions from this: (1) This indicates that if molecular structure has no effect on the initial cell state, there will not be a positive pairing between the molecular structure and morphological patterns captured by phenomics, making it indistinguishable from a control image. (2) There is an underlying assumption that the initial cell state \(x^0_i\) is constant. In accordance with this assumption we utilize experiments with a fixed cell line, \textit{HUVEC-19}, and a constant genetic background. Future works can relax this assumption by taking into account phenomics experiments of the cells prior to the perturbation. This can allow the models to generalize beyond a single cell line and to diverse genetic backgrounds. 

\section{Dataset}\label{sc:dataset}

Models have been trained using our in house training set and we have conducted our evaluation on two novel datasets and an open-source molecule dataset \cite{rxrx3}:
\begin{itemize}[leftmargin=*]
 \item \texttt{Training Set}: Our training dataset comprises 1,316,283 pairs of molecules and concentration concentration combinations, complemented by fluorescent microscopy images generated through over 2,150,000 phenomic experiments. 
 \item \texttt{Evaluation set 1 - Unseen Images + Seen Molecules}: The first set consists of unseen images and seen molecules. Unseen microscopy images are associated with 15,058 pairs of molecules and concentrations from the training set and selected randomly.
 \item \texttt{Evaluation set 2 - Unseen Images + Unseen Molecules}: The second set includes previously unseen molecules, and images (consisting of 45,771 molecule and concentration pairs). Predicting molecular identities of previously unseen molecular perturbations corresponds to zero-shot prediction. Scaffold splitting was used to split this validation dataset from training ensuring minimal information leakage. 
 \item \texttt{Evaluation set 3 - Unseen Dataset}: Finally, we utilize the RXRX3 dataset \cite{rxrx3}, an open-source out of distribution (OOD) dataset consisting of 6,549 novel molecule and concentration pairs associated with phenomic experiments. The distribution of molecular structures differs from previous datasets, making this a challenging zero-shot prediction task.
\end{itemize}

\subsection{Concentration Details}
Additional details regarding the number of molecules at significant concentrations of each evaluation set are available in Table \ref{tab:concentration_detail}.

\begin{table}[H]
    \centering
    \caption{Separated number of molecules for different concentrations at various pvalue cut-offs}
    \begin{adjustbox}{max width=\textwidth}
    \begin{tabular}{cccccccccc}
        \toprule
         &  \multicolumn{3}{c}{pvalue=1.0}  & \multicolumn{3}{c}{pvalue=.1}  & \multicolumn{3}{c}{pvalue=.01}\\

         \cmidrule(lr){2-4} \cmidrule(lr){5-7} \cmidrule(lr){8-10}
         
          {Concentration} & {Unseen Im.} & \thead{Unseen  \\ Im. + Mol.}  & {Unseen Data} & Unseen Im. & \thead{Unseen  \\ Im. + Mol.}   & Unseen Data & Unseen Im. & \thead{Unseen  \\ Im. + Mol.}  & Unseen Data\\

    % {Loss} & {Unseen Im.} &  \thead{{Unseen} \\ {Im. + Mol.}} & {Unseen Data} & {Unseen Im.} &  \thead{{Unseen} \\ {Im. + Mol.}} & {Unseen Data}\\
    % \midrule
         \midrule
         .1 & 1497 & 1109 & 0          & 387 & 170 & 0                 & 161 & 68 & 0\\
         .25 & 1775 & 1111 & 1638      & 600 & 203 & 237               & 334 & 121  & 165\\
         1.0 & 2721 & 11392 & 1639      & 1259 & 734 & 390              & 672 & 390  & 268\\
         2.5 & 1787 & 4018 & 1636       & 1329 & 644 & 516              & 929 & 413  & 375\\
         3.0 & 74 & 10454 & 0           & 12 & 1540 & 0                 & 4 & 729  & 0\\
         5.0 & 3 & 50 & 0               & 0 & 27 & 0                    & 0 & 20 & 0 \\
         1.0 & 2712 & 11392 & 1636     & 2544 & 8117 & 792             & 2116 & 4815 & 625 \\
         25.0 & 0 & 2916 & 0            & 0 & 1734 & 0                  & 0 & 950 & 0 \\
         \thead{ Unique \\ molecules} & 3026 & 14256 & 1639      & 2729 &  9857 & 823            & 2309 & 5778 & 642\\
         \bottomrule
    \end{tabular}
    \end{adjustbox}
    \label{tab:concentration_detail}

\end{table}

\section{Implementation Details}
\label{sc:appa}

In our experiments we report the top 1\% recall metric as it is agnostic to the size of the dataset used. Across different datasets, top 1 metric can correspond to varying levels of difficulty due to the number of negatives evaluated. Top 1\% can be used to compare models with different batch sizes, datasets, and evaluations with different number of concentrations.

\subsection{Hyperparameters}
Our design choices and utilized hyperparameters for is presented in Table \ref{tab:parameters}.
We set batch size to 512 through experiments presented in top section of Table \ref{tab:ScratchVsPhenom1} and Figure \ref{fig:scratchVsPhenom1} since training CLOOME model on images is not efficient compared to using pretrained models. In addition, results presented at bottom section of Table \ref{tab:ScratchVsPhenom1} are based on the best parameters found through described ablation studies (section \ref{sc:ablation}). 

\begin{table}[H]
    \centering
    \caption{Hyperparameter values utilized in our proposed MolPhenix training framework.}
    \begin{adjustbox}{max width=.75\textwidth}
    \begin{tabular}{|c|c|}
         \hline
         Parameter & Value\\
         \hline
         number of seeds & 3\\
         learning rate & 1e-3 \\
         weight decay & 3e-3 \\
         optimizer & \texttt{AdamW} \\
         training batch size & 8192 \\
         validation batch size & 12000 \\
         embedding dim & 512 \\
         model size & medium (38.7 M) \\
         model structure & 6 ResNet Blocks + 1 Linear layer + 1 ResNet Block + 1 Linear layer \\
         epochs & 100 \\
         self similarity clip val & .75 \\
         learnable temperature initialization & 2.302 \\
         learnable bias initialization & -1.0 \\
         Distance function & arctangent of l2 distance \\
         % $\alpha$ & .75 \\
         % $\beta$ & 1.25 \\
         \hline         
    \end{tabular}
    \end{adjustbox}
    \label{tab:parameters}
\end{table}

\subsection{Resource Computation}
\label{sc:compute}
We utilized an NVIDIA A100 GPU to train Molphenix using Phenom1 and MolGPS embeddings, which takes approximately $\sim$4.75 hours each. For loss comparison experiments, we run each model using 3 different seeds and 8 different losses, resulting in a total of 114 hours of GPU processing time. For concentration experiments we train 7 runs, one for each concentration, with 3 seeds each totaling 21 runs per set of parameters. With 25 sets of parameters evaluated (\ref{tab:appendix_whitin_dose_results_all_mols}), that amounts to 2,500 A100 compute hours. Moreover, we employed 8 NVIDIA A100 GPUs to train CLOOME model on phenomics images, with an average of 40 hour usage per run. Across three seeds, that amounts to $\sim$ 1000 hours of A100 GPU usage (8 GPUs for 40 hours 3 times).

Note that, without accounting for the time to train Phenom1, MolPhenix is 8.4 $\times$ faster than the CLOOME baseline.

\subsection{S2L Distance function}\label{sec:s2l_distance_appendix}
To calculate inter sample distances, we utilize arctangent of l2 distances between Phenom1 embeddings. More specifically, we calculate distances with

\begin{equation}
    \text{arctan}(\lVert z_{\mathbf{x}_i} - z_{\mathbf{x}_j} \rVert^2_2 / c) * \frac{4}{\pi} - 1,
\end{equation}

where \(c\) is a constant indicating the median l2 distance between a null set of embeddings. Empirically, we’ve found that setting similarities below a threshold \(k\) to 0 improves model performance: \(\ceil{w}^k\).

\begin{figure}[t]
    \centering
    \includegraphics[width=.65\textwidth]{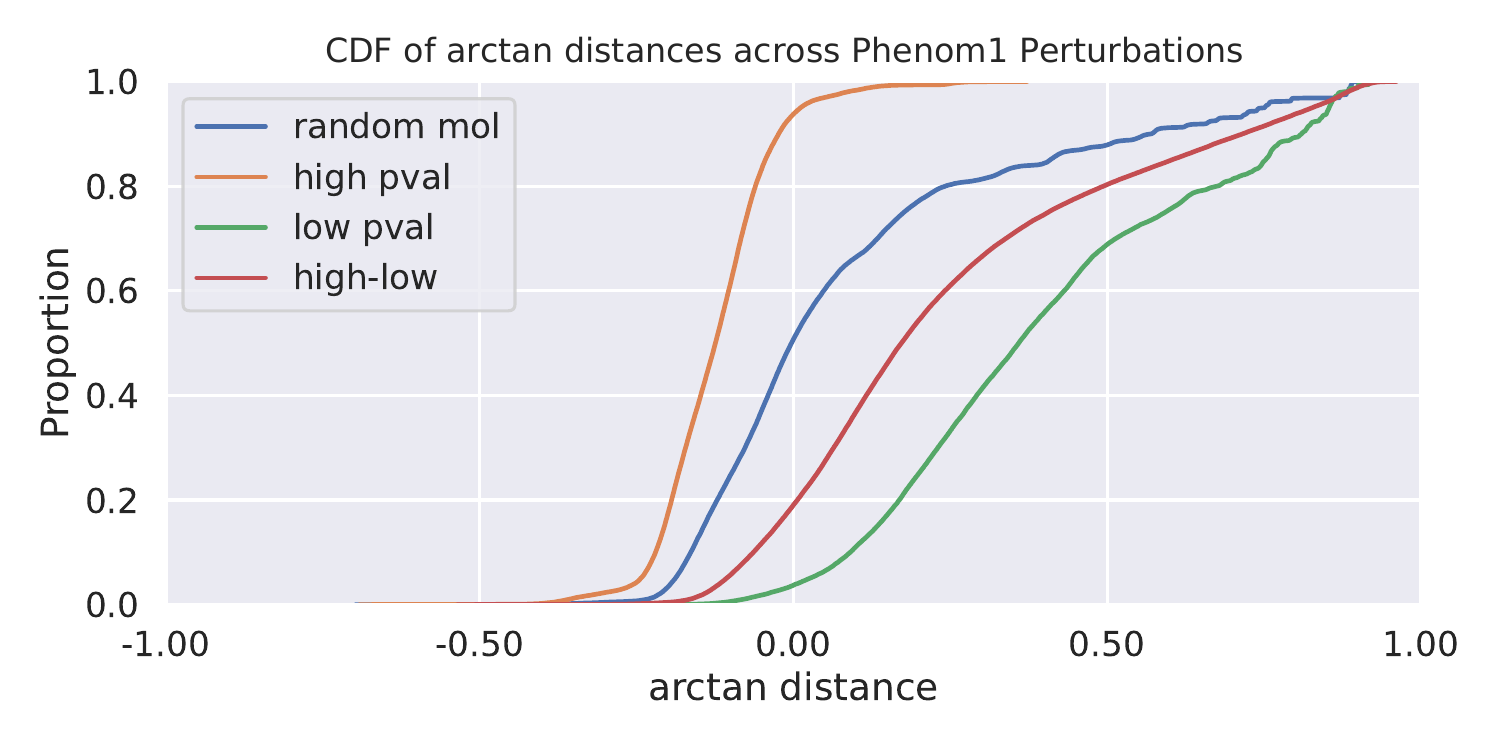}
    \\
    \includegraphics[width=.65\textwidth]{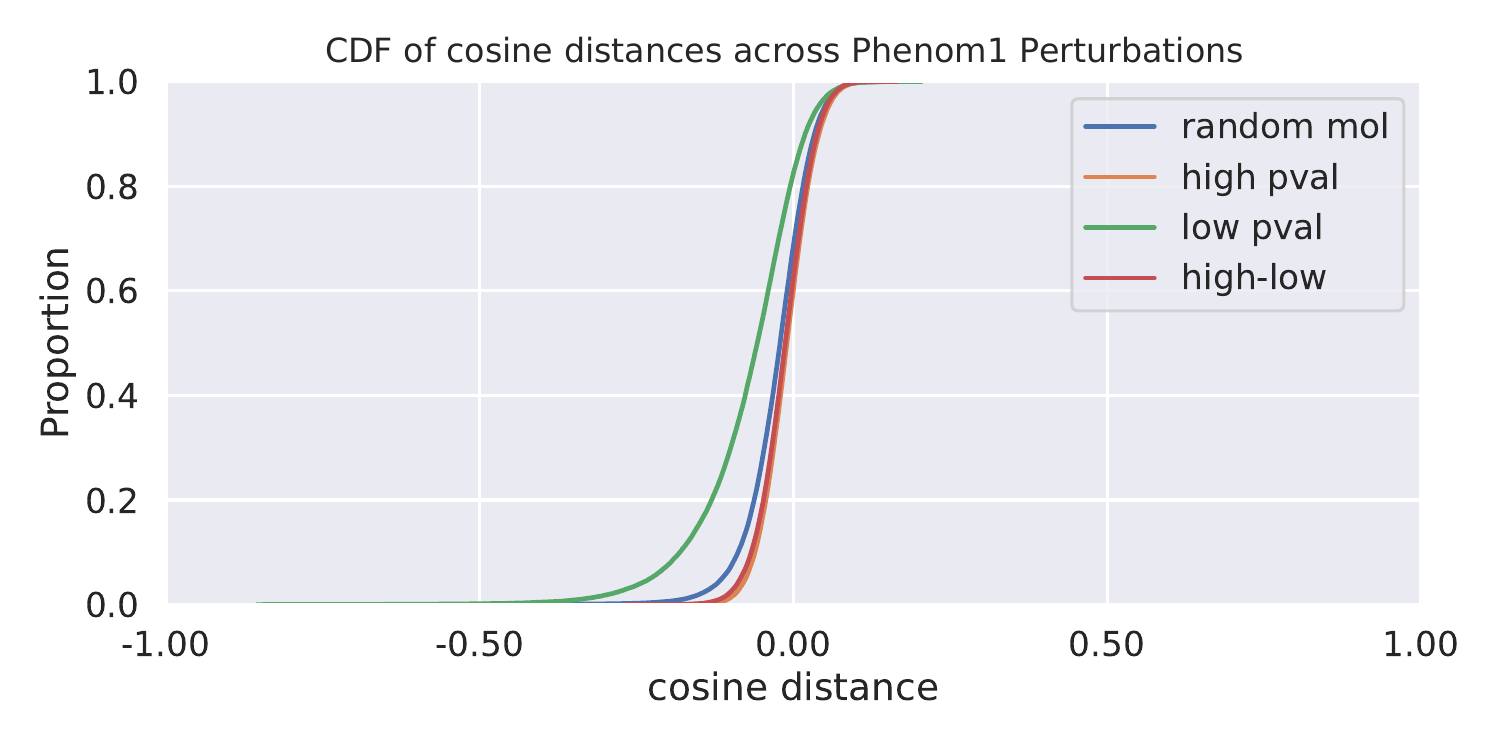}
    \caption{Plotted are cumulative densities of distance metrics for cosine similarity and arctangent of l2 distances between embeddings. Random mol corresponds to Phenom1 distances between random molecules, high pval corresponds to distances between molecules with high p-values, low pval corresponds to distances between active molecules with low p-values, finally high-low corresponds to distances between active and inactive molecules.}
    \label{fig:appendix_distance_Phenom1}
\end{figure}

Usage of arctan-l2 distances is motivated by an observation that cosine similarities do not effectively separate inactive molecules from other molecular pairs (Figure \ref{fig:appendix_distance_Phenom1}). To alleviate inactive molecule challenge, we require significant separation of CDF curves of inactive perturbations (p value > .9) and active molecules (p < .01). We observe that in both the plots using arctangent and cosine similarities achieves this purpose. However, if we compare high p-value curves with high-low, we find that in the case of cosine similarities they are almost identical. This indicates that the distribution of cosine similarities between active - inactive molecules is almost identical to that of inactive - inactive molecules. In contrast, when using arctangent similarities, we observe that the two CDF curves are well separated.

This property of l2 distances can inform our model training to identify inactive-inactive molecules. These results informed our decision to utilize arctangent of l2 distances to specify sample similarities for the \texttt{S2L} loss. 

% \vspace{-13cm}

\section{Additional Results} 
\label{sc:additional_results}

\subsection{Predicting molecular activity} \label{sec:molecular_activity}

\begin{figure}[H]
    \centering
    \includegraphics[width=.7\textwidth]{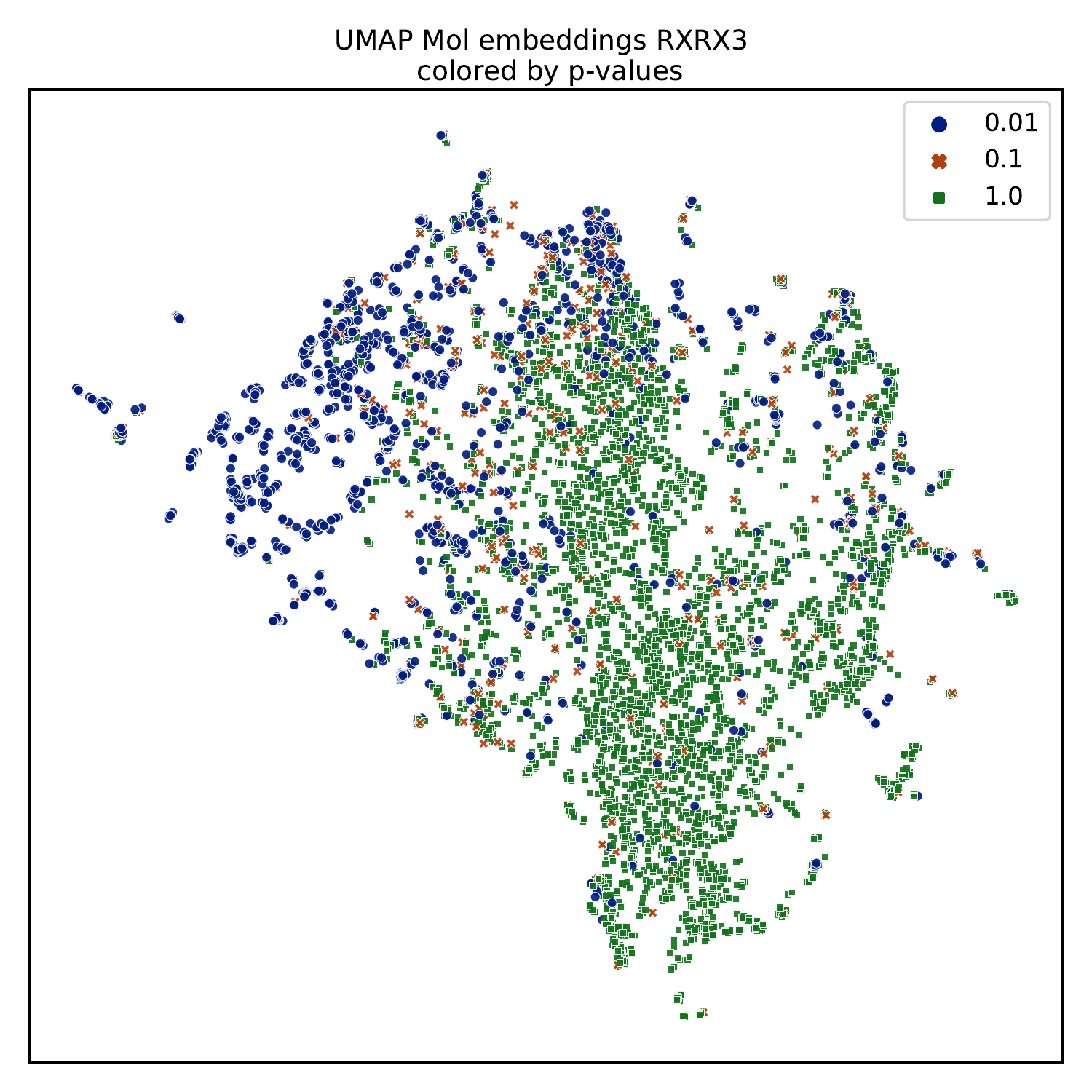}
    \includegraphics[width=.7\textwidth]{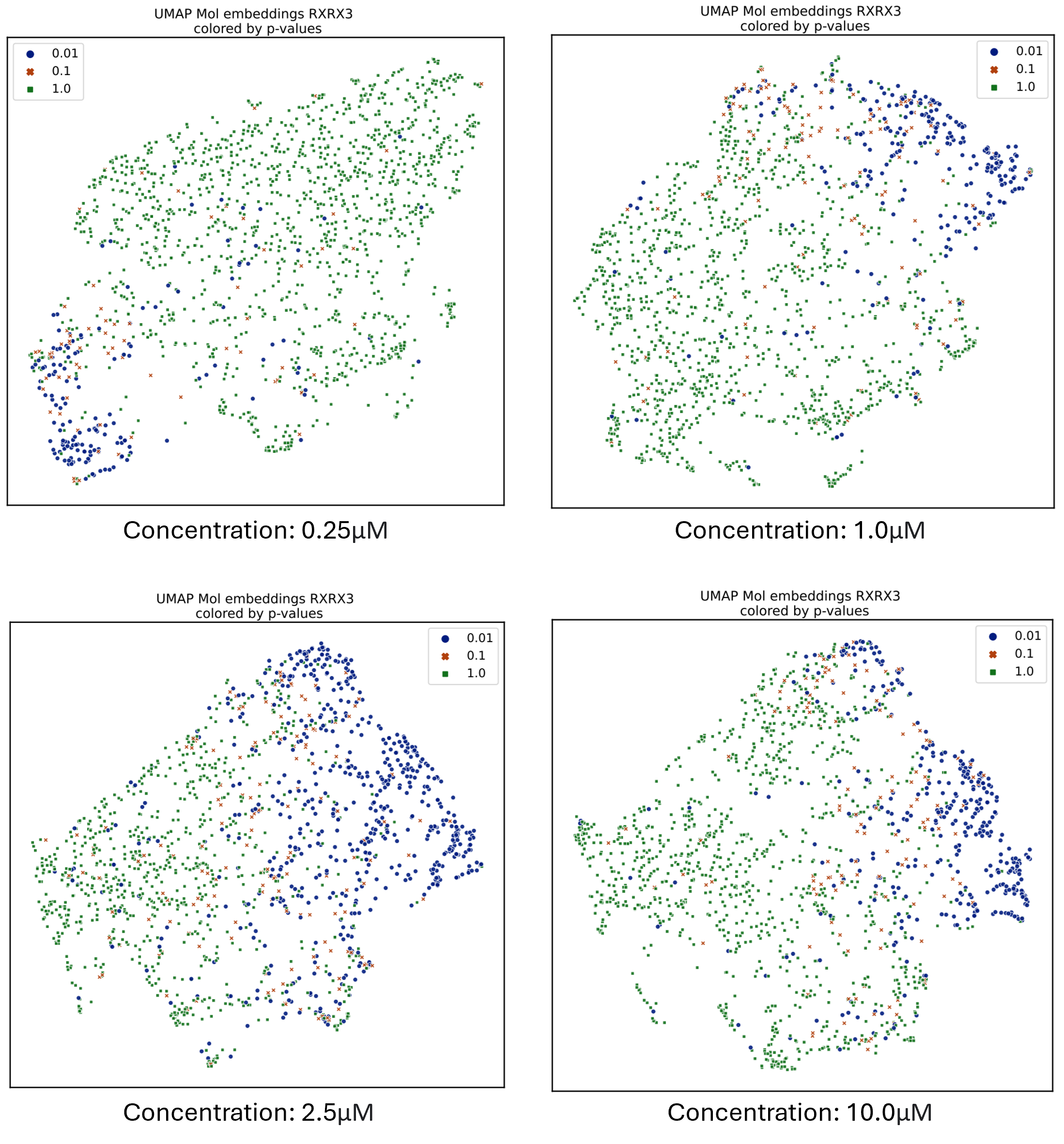}
    \caption{U-map demonstrating dimensionality reduction of the chemical embeddings of unseen dataset RXRX3. First two dimensions are visualized and points are colored corresponding to their activity measured in phenomics experiments. Activity is evaluated using p-values calculated using technical replicability of Phenom1 embeddings. Top plot shows the u-map figure of all chemical embeddings, and bottom figure contains u-map figure of representations at specific concentrations.}
    \label{fig:umap}
\end{figure}

\begin{figure}[h]
    \centering
    \includegraphics[width=.45\textwidth]{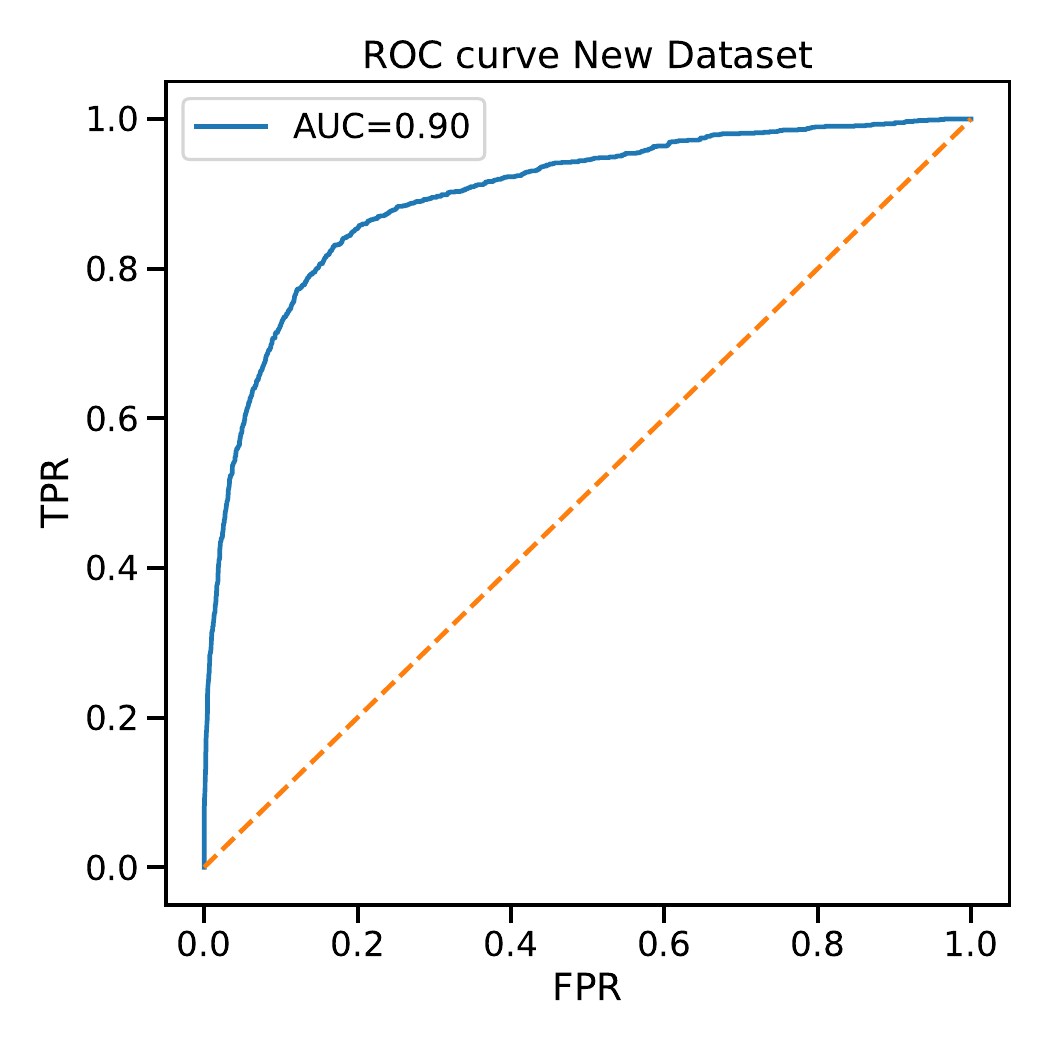}
    \includegraphics[width=.45\textwidth]{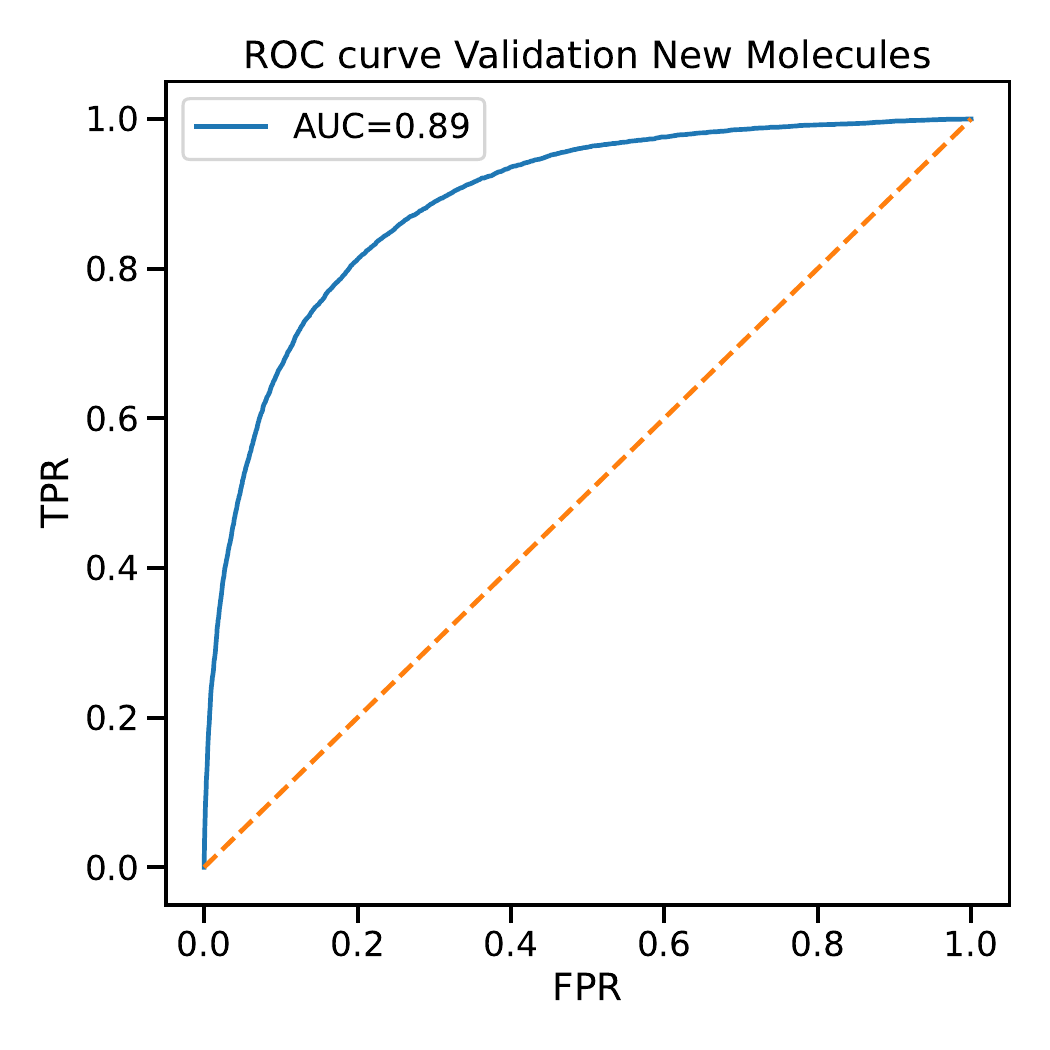} 
    \\
    \caption{\textbf{Top left:} ROC AUC of logistic regression predicting molecular activity on new dataset. \textbf{Top right:} ROC AUC of logistic regression predicting molecular activity on validation dataset with new molecules and new images.}
    % \textbf{Bottom:} U-map demonstrating dimensionality reduction of the chemical embeddings of unseen dataset RXRX3. First two dimensions are visualized and points are colored corresponding to their activity measured in phenomics experiments. Activity is evaluated using p-values calculated using technical replicability of Phenom1 embeddings.}
    \label{fig:predict_molecular_activity}
\end{figure}

Given the significance of identifying active molecules, we evaluate the ability of the chemical encoder to predict molecular activity. To do so, we assessed whether embeddings generated from the chemical encoder can be used to predict calculated p-values for unseen molecules. We fit a logistic regression on molecular embeddings from the training set, classifying whether a molecular perturbation and concentration have a p-value below .01. We find that the trained logistic regression is capable of predicting molecular activity on two downstream datasets with a non-overlapping set of molecules, Figure \ref{fig:predict_molecular_activity}. In addition, we provide a u-map of molecular embedding for the unseen dataset RXRX3, colored by p-value. We qualitatively observe a clustering of active molecules using a U-map (Figure \ref{fig:umap}). It demonstrates that predicting compounds activity is possible using MolPhenix chemical encoder as molecules representations are distinct, independent of the experimented dosage concentration.  
\newpage

\subsection{Zero Shot Biological Validation}\label{sec:zero_shot_validaiton}

We conduct a preliminary investigation into whether MolPhenix can be used to identify biological relationships without the need for conducting the underlying experiments. To this end, we evaluate on a subset of ChEMBL with curated pairs of gene knockouts and molecular perturbants \cite{mendez2019chembl}. These pairs of perturbations were curated due to the similarity of their effects on cells, although these might not always be captured through phenomic experiments. Thus, there is maximum performance that can be reached through just phenomic data, which we assume to be achieved by experimental data embedded using Phenom1. 

To evaluate MolPhenix's ability to identify previously known biological associations directly from data, we embed phenomics experiments from gene knockouts using the vision encoder. To perform in-silico screening, we then embed the molecular structures associated with positive pairs using the chemical encoder. Generating molecular embeddings and the corresponding concentrations does not utilize any experimental data. We then calculate cosine similarities between embeddings of phenomics experiments evaluating gene knockouts, and representations of the chemical representations along with encoded concentrations. Using the computed cosine similarities we are then able to assess whether MolPhenix is capable of identifying known associations between gene knockouts and molecular structures. Since there is no information on molecular concentration at which the cells must be treated with, we repeat the experiment across 4 concentrations. To get a null distribution of cosine similarities we take pairs of genes knockouts and molecules for which there are no annotated relationships. We calculate a cut-off for a low and high percentiles, and then evaluate what percentage of pairs of genes and molecules with known relationships exceed the set thresholds. 

\begin{figure}[H]
    \centering
    \includegraphics[width=\textwidth]{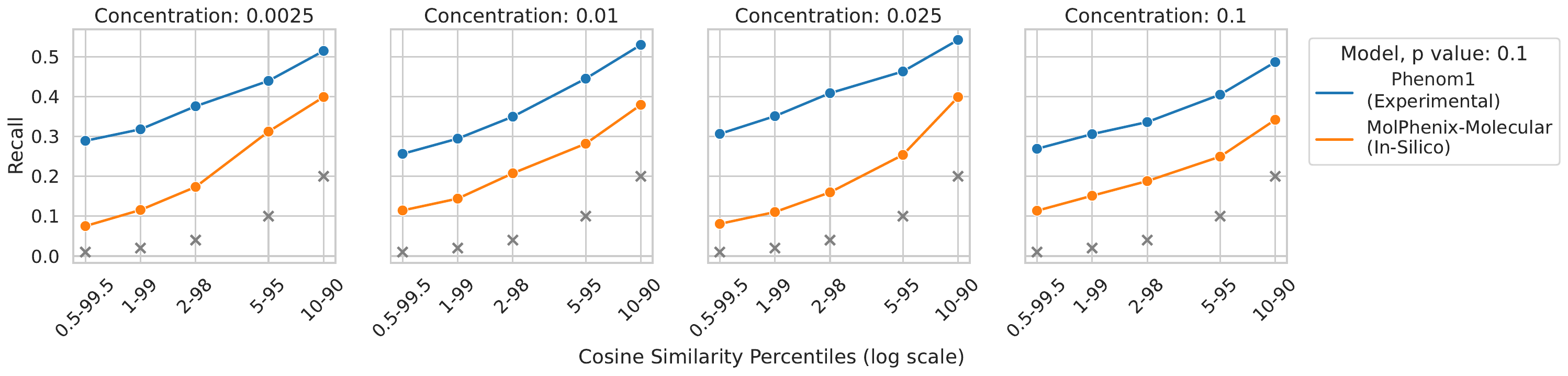}
    \caption{Evaluation of 0-shot ChEMBL identification of gene knockout and molecular phenomic similarities. On the X axis are percentile ranges, at which points the threshold is computed for cosine similarities. On the y axis is plotted total recall of recovered known interactions. Grey \textit{x} plotted for each range indicate baseline recall. Orange line indicates MolPhenix-Molecular encoding of chemical compounds and MolPhenix-Vision for encoding gene knockout phenomics experiment. Blue line indicates Phenom1 encoding of phenomics experiments for both the molecular perturbation and gene knockouts. In-silico encoding of molecular perturbation, as well as the corresponding concentration, recovers a significant fraction of observed interactions.}
    \label{fig:chembl_results}
\end{figure}

Figure \ref{fig:chembl_results} demonstrates that in-silico screening using MolPhenix Molecular encoder is capable of recovering a significant portion of known interactions. This is performed without the use of experimental data on the molecular encoder. It is difficult to estimate an upper bound on the expected performance due to uncertainty in the quality of curation of known pairs, presence of unknown associations between genes and molecules, and uncertainty regarding molecular concentration. There is a clear trend however that MolPhenix molecular encoder is capable of recovering a meaningful fraction of these interactions. 

\subsection{Molecular Property Prediction}\label{sec:tdc_polaris}

We expand our evaluation with additional experiments supporting the utility of MolPhenix beyond retrieval. We conduct a KNN evaluation of the MolPhenix latent space, assessing the learned embedding on 35 molecular property prediction tasks across the Polaris and TDC datasets (Table \ref{fig:tdc_polaris_fps} and \ref{fig:tdc_polaris_molgps}). We find that MolPhenix trained with fingerprint embeddings consistently outperforms standalone input fingerprints, demonstrating that the MolPhenix latent space effectively clusters molecules according to their biological properties. We observed an interesting effect where prediction quality is positively correlated with implied dosage, indicating that MolPhenix learns dosage-specific effects. In addition, utilizing 

\begin{table}[H]
    \centering
    \caption{Comparison of a KNN applied on MolPhenix molecular embedding with \textbf{traditional fingerprints} on different tasks of TDC and Polaris datasets. Mean results for TDC, Polaris and together are available in the last three columns. Binary fingerprints use tanimoto similarity, while floating-point fingerprints use cosine similarity.}
    \includegraphics[width=1\linewidth]{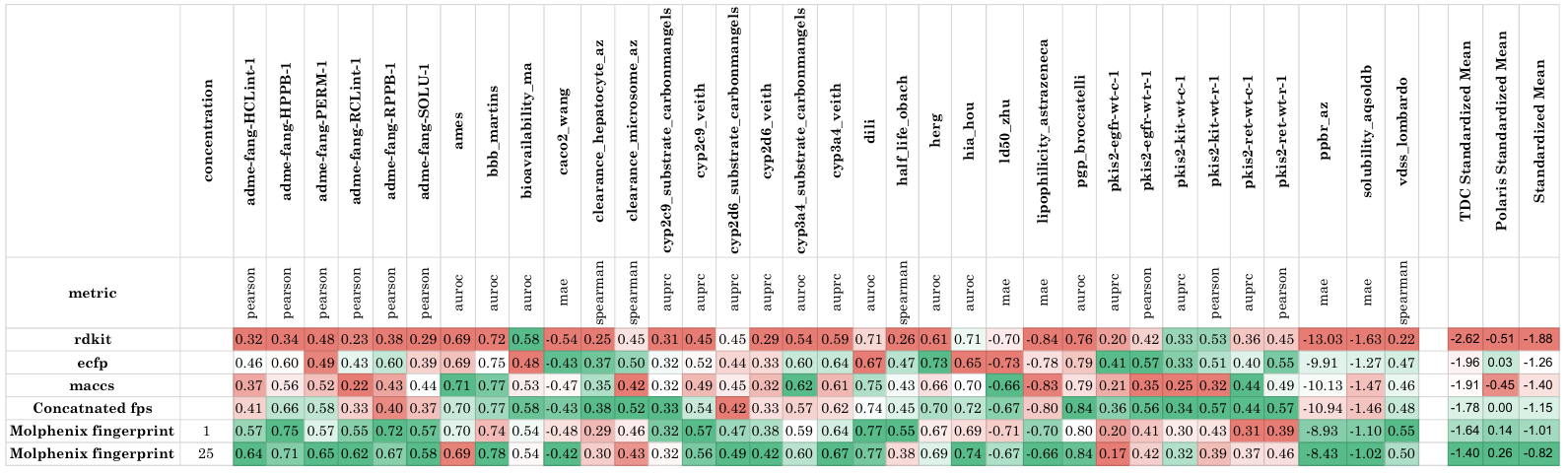}
    
    \label{fig:tdc_polaris_fps}
\end{table}

\begin{table}[H]
    \centering
    \caption{Comparison of a KNN applied on MolPhenix molecular embedding with \textbf{MolGPS} on different tasks of TDC and Polaris datasets. Mean results for TDC, Polaris and together are available in the last three columns.}
    \includegraphics[width=1\linewidth]{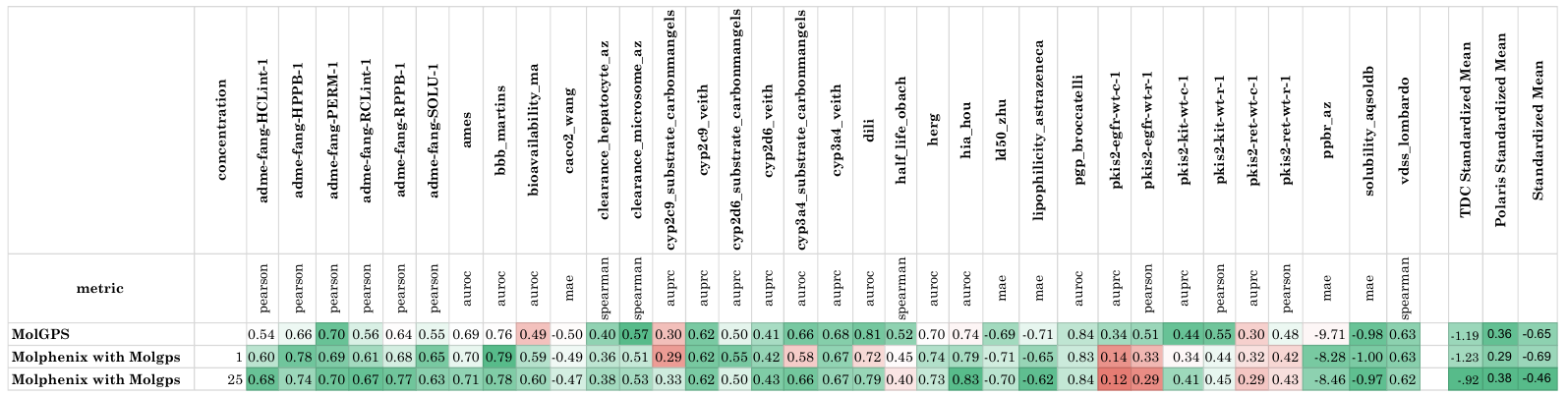}
    
    \label{fig:tdc_polaris_molgps}
\end{table}

\subsection{Addressing Challenges in Contrastive Phenomic Retrieval}

Table \ref{tab:appendix_cross_dose_results} and \ref{tab:appendix_cross_dose_results_all_mols} show the complete Top 1\% and 5\% results of evaluation on cumulative concentrations on only active and all molecules, respectively. Similarly, Table \ref{tab:appendix_within_dose} and \ref{tab:appendix_whitin_dose_results_all_mols} presents the full retrieval results of held-out concentrations experiments.
% While other loss functions may only show acceptable performance in only one category, 
In comparison to prior loss functions, S2L loss objective demonstrates consistent high retrieval rate in all tasks and molecular groups (i.e. all or active molecules), while using the same modality (Phenom1) and with or without explicit concentration information.

\begin{table}
    \centering
     \caption{\textbf{Evaluation on cumulative concentrations for active molecules:} Average Top-1\% and Top-5\% recall accuracies of methods utilizing different contrastive learning loss functions and concentration encoding information. We evaluate all methods on unseen images, unseen images and unseen molecules and an unseen dataset for zero-shot retrieval. Entries in \textbf{bold} denote best performance when the loss function is fixed while entries in \colorbox{babyblue}{highlight} denote best performance across all guidelines.}
  \begin{adjustbox}{max width=\textwidth}
    \begin{tabular}{ccccccccccc}
    % \toprule
        \toprule
        &   &  & \multicolumn{4}{c}{\textbf{top-1\%}} & \multicolumn{4}{c}{\textbf{top-5\%}} \\
         \cmidrule(lr){4-7} \cmidrule(lr){8-11}
        Method  & Explicit  & Modality & Unseen Images & Unseen Images +& Unseen Dataset & Avg. & Unseen Images & Unseen Images +& Unseen Dataset & Avg. \\
         & Concentration (ours) &  &  & Unseen Molecules & (0-shot) & &  & Unseen Molecules & (0-shot) & \\
        \toprule
        % CLIP & \xmark & Images & - & - & - & - & - & - & -  \\
        CLIP & \xmark & Phenom1  & \textbf{.3373} & \textbf{.4228} & \textbf{.1514} & .3038 & \textbf{.6162} & \textbf{.7182} &  \textbf{.3660} & .5668 \\
        % CLIP & \xmark & Phenom1  & \textbf{.4693} & \textbf{.4465} & \textbf{.1951} & .3703 & \textbf{.7956} & \textbf{.7363}	&  \textbf{.4339} & .6552 \\
        \midrule
        % Hopfield-CLIP  & Images & - & - & - & - & - & - & - & - \\
        Hopfield-CLIP  & \xmark & Phenom1  & \textbf{.2578} & \textbf{.3559} & \textbf{.1256} & .2464 & \textbf{.5457} & \textbf{.6751}	& \textbf{.3270} & .5159 \\
        % Hopfield-CLIP  & \xmark & Phenom1  & \textbf{.3616} & \textbf{.3946} & \textbf{.1407} & .2989 & \textbf{.6963} & \textbf{.7039}	& \textbf{.3635} & .5879 \\
        \midrule
        % InfoLOOB  & \xmark & Images & - & - & - & - & - & - & - & - \\
        InfoLOOB  & \xmark & Phenom1  & \textbf{.3351} & \textbf{.4206} & \textbf{.1563} & .3040 & \textbf{.6128} & \textbf{.7204}	&  \textbf{.3730} & .5687 \\
        % InfoLOOB  & \xmark & Phenom1  & \textbf{.4687} & \textbf{.4413} & \textbf{.2072} & .3724 & \textbf{.7902} & \textbf{.7343}	&  \textbf{.4420} & .6555 \\
        \midrule
        % CLOOME & \xmark & Images & - & - & - & - & - & - & - & -\\
        CLOOME & \xmark & Phenom1  & .3572 & .4348 & .1658 & .3193 & .6330 & .7259 & .3918 & .5836 \\
        CLOOME & sigmoid & Phenom1  & .5813 & .4968 & .2360 & .4380	& .8748 & .7658 & .4859 & .7088 \\
        CLOOME & logarithm & Phenom1  & \textbf{.6057} & \textbf{.5255} & \textbf{.2445}	& \textbf{.4586}	& \textbf{.8858}	& .8117	& \textbf{.4957}	& \textbf{.7310} \\
        CLOOME & one-hot & Phenom1  & .5967 & .5255 & .2380 & .4534 & .8800 & \textbf{.8120} & .4829 & .7250 \\
        
        % CLOOME & \xmark & Phenom1  & \textbf{.4500} & \textbf{.4532} & \textbf{.1927} & .3653 & \textbf{.7669} & \textbf{.7317} & \textbf{.4406} & .6464 \\
        \midrule
        DCL & \xmark & Phenom1 & .6363 & .6177 & .3184 & .5241 & .8638 & .8180 & .5632 & .7483 \\
        % DCL & \cmark & \xmark & Phenom1  & .8497 & .6319 & .3986 & .6267 &  .9563 & .8311 & .6527 & .8133 \\
        DCL & sigmoid & Phenom1  & .8858 & .6694 & .4527 & .6693 & \textbf{.9600} & .8472 & .6845 & .8305 \\
        DCL & logarithm & Phenom1 & .8934 & .6952 & .4511 & .6799 & .9581 & \textbf{.8788} & \textbf{.6889} & \textbf{.8419}\\
        DCL & one-hot & Phenom1 & \textbf{.8901} & \textbf{.7002} & \textbf{.4601} & \textbf{.6834} & .9591 & .8770 & .6873 & .8411 \\
        
        \midrule
        CWCL & \xmark & Phenom1  & .7091 & .6529 & .3556 & .5725 & .9018 & .8368 & .6027 & .7804 \\
        % CWCL & \cmark & \xmark & Phenom1  & .8867 & .6585 & .4185 & .6545 & .9673 & .8482 & .6638 & .8264 \\
        CWCL & sigmoid & Phenom1 & .9138 & .6985 & .4810 & .6977 & .9681 & .8643	& .7070 & .8464 \\
        CWCL & logarithm & Phenom1 & .9141 & .7248 & .4815 & .7068 & .9651 & .8920 & \textbf{.7131} & \textbf{.8567} \\
        CWCL & one-hot & Phenom1 & \textbf{.9128} & \textbf{.7261} & \textbf{.4850} & \textbf{.7079} &\textbf{.9665} & \textbf{.8927} & .6998 & .8530\\
        \midrule
        SigLip & \xmark & Phenom1  & .7763 & .6401 & .3396 & .5853 & .9361 & .83038 & .5714 & .7792 \\
        % SigLip & \cmark & \xmark & Phenom1  & .9215 & .6580 & .4104 & .6633  & .9783 & .8412 & .6341 & .8178 \\
        SigLip & sigmoid & Phenom1  & .9463 & .6931 & .4576 & .6990 & .9816 & .8606 & .6759 & .8393 \\
        SigLip & logarithm & Phenom1  & \colorbox{babyblue}{\textbf{.9493}} & .7256 & .4868 & \textbf{.7205} & .9814  & \textbf{.8926} & .7019 & \textbf{.8586}\\
        SigLip & one-hot & Phenom1  & .9489 & \textbf{.7210} & \textbf{.4859} & .7186 & \textbf{.9823} & .8868 & \textbf{.7045} & .8578 \\
        \midrule
        % MolPhenix (ours) & \xmark & \xmark & Phenom1 & .7350 & .6509 & .3333 & .5730 & .9198 & .8395 & .5572 & .7721 \\
        MolPhenix (ours) & \xmark & Phenom1  & .9097 & .6759 & .4181 & .6679 & .9768 & .8539 & .6436 & .8247 \\
        MolPhenix (ours) & sigmoid & Phenom1  & .9423 & .7155 & .4573 & .7050 & .9808 & .8775 & .6778 & .8453 \\
        MolPhenix (ours) & logarithm & Phenom1  & .9426 & .7451 & .4727 & .7201 & .9808 & .8964 & .6952 & .8574 \\
        MolPhenix (ours) & one-hot & Phenom1  & \textbf{.9430} & \textbf{.7490} & \textbf{.4850} & \textbf{.7256} & \textbf{.9816} & \textbf{.8984} & \textbf{.7040} & \textbf{.8613} \\
        \midrule 
        MolPhenix (ours) & \xmark &Phenom1 + MolGPS  & .9105 & .6710 & .4501 & .6772 & .9755 & .8527 & .7098 & .8460 \\
        MolPhenix (ours) & sigmoid &Phenom1 + MolGPS & .9395 & .7034 & .5252 & .7227 & .9811 &	.8729 & .7630 & .8723 \\
        MolPhenix (ours) & logarithm &Phenom1 + MolGPS & .9413 & .7505 & .5473 & .7463 & .9811 & \colorbox{babyblue}{\textbf{.9085}} & \colorbox{babyblue}{\textbf{.7878}} & \colorbox{babyblue}{\textbf{.8924}} \\
        MolPhenix (ours) & one-hot &Phenom1 + MolGPS & \textbf{.9430} & \colorbox{babyblue}{\textbf{.7514}}  & \colorbox{babyblue}{\textbf{.5577}} & \colorbox{babyblue}{\textbf{.7507}} &  \colorbox{babyblue}{\textbf{.9830}} & .9043 & .7821 & .8898 \\
        \bottomrule
    \end{tabular}
    \end{adjustbox}
    \label{tab:appendix_cross_dose_results}
\end{table}

% cwcl better in 1% of nothing, equal in 1% of all cumilative, worse in rest
% siglip equal in 1% of active cumilative and helout, worse in top 5%. worse in top %1 ALL MOL
%s2l best in top 5 all. equal top 1 active both, best in top 1 all mol
\begin{table}
    \centering
    \caption{\textbf{Evaluation on held-out concentration for active molecules:} Average Top-1\% and Top-5\% recall accuracies of methods utilizing different contrastive learning loss functions and concentration encoding information. We evaluate all methods on unseen images, unseen images and unseen molecules and an unseen dataset for zero-shot retrieval. Entries in \textbf{bold} denote highest performance when the loss function is fixed while entries in \colorbox{babyblue}{highlight} denote highest performance across all guidelines.}
  \begin{adjustbox}{max width=\textwidth}
    \begin{tabular}{ccccccccccc}
        \toprule
        &    &  & \multicolumn{4}{c}{\textbf{top-1\%}} & \multicolumn{4}{c}{\textbf{top-5\%}} \\
        \cmidrule(lr){4-7} \cmidrule(lr){8-11}
        Method  & Explicit & Modality & Unseen Images & Unseen Images +& Unseen Dataset & Avg. & Unseen Images & Unseen Images +& Unseen Dataset & Avg. \\
         & Concentration (ours) &  &  & Unseen Molecules & (0-shot) & &  & Unseen Molecules & (0-shot) & \\
        \toprule
        % CLIP & \xmark & Images & - & - & - & - & - & - \\
        CLIP & \xmark & Phenom1  & \textbf{.2109} & \textbf{.2425} & \textbf{.1519} & \textbf{.2018} & \textbf{.4458} & \textbf{.4968} & \textbf{.3591} & \textbf{.4339} \\
        % CLIP & \xmark & Phenom1  & \textbf{.3273} & \textbf{.2610} & \textbf{.1865} & .2582 & .6407 & .5254 & .3985 & .5215 \\
        \midrule
        % Hopfield-CLIP & \xmark & Images & - & - & - & - & - & - \\
        Hopfield-CLIP & \xmark & Phenom1 & \textbf{.1581} & \textbf{.2034} & \textbf{.1198} & \textbf{.1604} & \textbf{.3783} & \textbf{.4413} & \textbf{.3045} & \textbf{.3747}\\
        % Hopfield-CLIP & \xmark & Phenom1 & \textbf{.2331} & \textbf{.2207} & \textbf{.1496} & .2011 & .5118 & .4746 & .3508 & .4457\\
        \midrule
        % InfoLOOB & \xmark & Images & - & - & - & - & - & - \\
        InfoLOOB & \xmark & Phenom1 & \textbf{.2122} & \textbf{.2496} & \textbf{.1501} & \textbf{.2040} & \textbf{.4443} & \textbf{.5003} & \textbf{.3515} & \textbf{.4320}\\
        % InfoLOOB & \xmark & Phenom1 & \textbf{.3290} & \textbf{.2585} & \textbf{.1818} & .2564 & .6391 & .5187 & .4023 & .5200\\
        \midrule
        % CLOOME & \xmark & Images & - & - & - & - & - & - \\
        CLOOME & \xmark & Phenom1 & .2164 & .2461 & .1479 & .2035 & .4590 & .4956 & .3528 & .4358 \\
        CLOOME & sigmoid & Phenom1 & .3338 & .2681	& .1801 & .2606 & .6037 & .5202	& .3879	& .5039 \\
        CLOOME & logarithm & Phenom1 & .3094 & .2345 & .1665 & .2368 & .5960 & .4874 & .3534	& .4790 \\
        CLOOME & one-hot & Phenom1 & .3073  & .2040	& .1670	& .2261 & .5997 & .4246	& .3657	& .4633 \\
        
        % CLOOME & \xmark & Phenom1 & \textbf{.2961} & \textbf{.2677} & \textbf{.1749} & .2462 & .5876 & .5185 & .3996 & .5019 \\
        \midrule
        DCL & \xmark & Phenom1 & .4717 & .4027 & .2841 & .3861 & .7352 & .6579 & .5322 & .6417\\
        % DCL & \cmark & \xmark & Phenom1 & \textbf{.7503} & \textbf{.4244} & \textbf{.3714} & .5153 & .9260 & .6631 & .6215 & .7368\\
        DCL & sigmoid & Phenom1 & \textbf{.7282} & \textbf{.4100} & \textbf{.3560}& \textbf{.4980 }& \textbf{.9226} & \textbf{.6561} & \textbf{.6015 }& \textbf{.7267} \\
        DCL & logarithm & Phenom1 & .6903 & .3558 & .3211 & .4557 & .8869 & .6146 & .5667 & .6894 \\
        DCL & one-hot & Phenom1 & .6562 & .3607 & .3272 & .4480 & .8831 & .5983 & .5659 & .6824 \\
        \midrule
        CWCL & \xmark & Phenom1 & .5731 & .4403 & .3232 & .4455 & .8218 & .6833 & .5706 & .6919\\
        % CWCL & \xmark & Phenom1 & \textbf{.7954} & \textbf{.4643} & \textbf{.3944} & .5513 & .9434 & .6957 & .6345 & .7578 \\
        CWCL & sigmoid & Phenom1 & \textbf{.7780} & \textbf{.4425} & \textbf{.3777} & \textbf{.5327} & \textbf{.9386}	& \textbf{.6844} & \textbf{.6244} & \textbf{.7491}\\
        CWCL & logarithm & Phenom1 & .7452 & .3989 & .3523 & .4988 & .9117 & .6482 & .5962 & .7187\\
        CWCL & one-hot & Phenom1 & .7048 & .4009 & .3593 &  .4883 & .9037 & .6284 & .6061 & .7127 \\
        \midrule
        SigLip & \xmark & Phenom1  & .5718 & .4217 & .3021 & .4318 & .8104 & .6602 & .5176 & .6627\\
        % SigLip & \xmark & Phenom1 & \colorbox{babyblue}{\textbf{.8395}} & .4590 & .3723 & .5569 & .9615 & .6876 & .6036 & .7509 \\
        SigLip & sigmoid & Phenom1 & \colorbox{babyblue}{\textbf{.8366}} & \textbf{.4640} & \textbf{.3830} & \textbf{.5612} & \textbf{.9623} & \textbf{.7023} & \textbf{.6080} & \textbf{.7575}\\
        SigLip & logarithm & Phenom1 & .8097 & .4391 & .3747 & .5411 & .9437 & .6746	& .6046 & .7409 \\
        SigLip & one-hot & Phenom1 & .7561 & .4020 & .3345 & .4975 & .9279 & .6248 & .5557 & .7028 \\
        \midrule
        % MolPhenix (ours) & \xmark & Phenom1 & .5942 & .4315 & .3129 & .4462 & .8328  & .6711 & .5357 & .6798 \\
        MolPhenix (ours) & \xmark & Phenom1 & \textbf{.8334} & .4615 & \textbf{.3792} & .5580 & .9638 & .6937	&  \textbf{.6128} & .7567 \\
        MolPhenix (ours) & sigmoid & Phenom1 & .8256 & \textbf{.4692} & .3765 & \textbf{.5571} & \colorbox{babyblue}{\textbf{.9638}} & \textbf{.7068} & .6115 & \textbf{.7607} \\
        MolPhenix (ours) & logarithm & Phenom1 & .7953 & .4466 & .3664 & .5361 & .9466 & .6889	& .5924 & .7426\\
        MolPhenix (ours) & one-hot & Phenom1 & .7489 & .4088 & .3379 & .4985 & .9325 & .6465 & .5644 & .7144 \\
        \midrule
        MolPhenix (ours) & \xmark &Phenom1 \& MolGPS & \textbf{.8277} & .4739 & .4071 & .5695 & \textbf{.9602} & .7041 & .6798 & .7813 \\
        MolPhenix (ours) & sigmoid &Phenom1 \& MolGPS & .8218 & \colorbox{babyblue}{\textbf{.4771}} & .4287 & \colorbox{babyblue}{\textbf{.5758}} & .9588 & .7117 & \colorbox{babyblue}{\textbf{.7045}} & \colorbox{babyblue}{\textbf{.7916}}\\
        MolPhenix (ours) & logarithm &Phenom1 \& MolGPS & .7836 & .4757 & \colorbox{babyblue}{\textbf{.4297}} & .563	& .9402 & \colorbox{babyblue}{\textbf{.7138}} & .7011 & .7850 \\
        MolPhenix (ours) & one-hot &Phenom1 \& MolGPS & .7391 & .4307  & .3940 & .5212 & .9198 & .6724 & .6698 &  .7540 \\
        \bottomrule
    \end{tabular}
    \end{adjustbox}
    \label{tab:appendix_within_dose}
\end{table}

\begin{table}
    \centering
        \caption{\textbf{Evaluation on cumulative concentrations for active and inactive perturbations} Average Top-1\% and Top-5\% Recall accuracy of methods utilizing different contrastive learning methods. Best performing methods are highlighted in \textbf{bold}.}
          \begin{adjustbox}{max width=\textwidth}
            \begin{tabular}{ccccccccccc}
                \toprule
                 &  &  & \multicolumn{4}{c}{\textbf{top-1\%}} & \multicolumn{4}{c}{\textbf{top-5\%}} \\
                \cmidrule(lr){4-7} \cmidrule(lr){8-11}
                 Loss & Explicit & Modality & Unseen Images & Unseen Images +& Unseen Dataset & Avg. & Unseen Images & Unseen Images +& Unseen Dataset & Avg. \\
                & Concentration &  &  & Unseen Molecules & (0-shot) & &  & Unseen Molecules & (0-shot) & \\
                \toprule
                CLIP &  \xmark & Phenom1  & .1761 & .1867 & .0734 & .1454 & .3710 & .3769 & .2065 & .3181 \\
                \midrule
                Hopfield-CLIP  & \xmark & Phenom1 & .1531 &	.1709 & .0673	& .1304 & .3464 & .3637 & .1942	& .3014 \\
                \midrule
                InfoLOOB & \xmark & Phenom1 & .1746 & .1860 & .0745 & .1450 & .3697 & .3756 & .2058 & .3170 \\
                \midrule
                CLOOME  & \xmark & Phenom1 &  .1968 & .2005 & .0911 & .1628 & .3938 & .3888 & .2321 & .3383 \\
                CLOOME & sigmoid & Phenom1 & .3875 & .2592 &	.1415 & .2627 & .5662 & .4601 &	.2940 & .4401 \\
                CLOOME & logarithm & Phenom1 & \textbf{.4088} & .3046 & \textbf{.1503} & .2879 & .5730 & .5166 & .3053 & .4650 \\
                CLOOME & one-hot & Phenom1 & .4080 & \textbf{.3123} & .1496 & \textbf{.2900} & \textbf{.5801} & \textbf{.5306} & \textbf{.3054} & \textbf{.4720} \\
                \midrule
                DCL & \xmark & Phenom1 & .3277 & .2562 & .1364 & .2401 & .4856 & .4170 & .2768 & .3931 \\ 
                DCL & sigmoid & Phenom1 & .4881 & .3380 & .2009 & .3423 & .6222 & .5186 & .3381 & .4930 \\ 
                DCL & logarithm & Phenom1 & .4983 &	.3615 & .2122 & .3573 & .6311 & .5581 & .3587 & .5160\\ 
                DCL & one-hot & Phenom1 & \textbf{.5226} & \textbf{.3790} & \textbf{.2288} & \textbf{.3768} & \textbf{.6791} & \textbf{.5870} & \textbf{.3968} & \textbf{.5543} \\ 
                \midrule
                CWCL  & \xmark & Phenom1 & .3635 & .2696 & .1526 & .2619 & .5122 & .4267 & .2933 & .4107\\
                CWCL  & sigmoid & Phenom1 &.5070 & .3457 & .2101 & .3542 & .6378 & .5272 & .3462 & .5037\\
                CWCL  & logarithm & Phenom1 & .5146 & .3725 & .2246 & .3706 & .6437 & .5733 & .3660 & .5277\\
                CWCL  & one-hot & Phenom1 & \textbf{.5401} & \colorbox{babyblue}{\textbf{.3849}} & \textbf{.2336} & \textbf{.3862} & \textbf{.6882} & \colorbox{babyblue}{\textbf{.5991}} & \textbf{.4001} & \textbf{.5625}\\
                \midrule
                SigLip & \xmark & Phenom1  & .3729 & .2544 & .1470 & .2581 & .5200 & .4179 & .2838 & .4072 \\
                SigLip & sigmoid & Phenom1  & .5021 & .3275 & .2072 & .3456 & .6360 & .5231 & .3430 & .5007 \\
                SigLip & logarithm & Phenom1  & .5156 & .3636 & .2233 & .3675 & .6452 & .5689 & .3653 & .5265 \\
                SigLip & one-hot & Phenom1  & \textbf{.5354} & \textbf{.3745} & \textbf{.2317} & \textbf{.3805} & \textbf{.6858} & \textbf{.5928} & \textbf{.3945} & \textbf{.5577} \\
                \midrule
                S2L (ours)  & \xmark & Phenom1  & .4688 & .2852 & .1838 & .3126 & .5970 & .4519 & .3171 & .4554 \\
                S2L (ours)  & sigmoid & Phenom1  & .5071 & .3441 & .2144 & .3552 & .6428 & .5315 & .3554 & .5099 \\
                S2L (ours)  & logarithm & Phenom1  & .5183 & .3700 & .2275 & .3720 & .6492 & .5650 & .3756 & .5300\\
                S2L (ours)  & one-hot & Phenom1  & \colorbox{babyblue}{\textbf{.5433}} & \textbf{.3819} & \textbf{.2384} & \textbf{.3879} & \colorbox{babyblue}{\textbf{.6954}} & \textbf{.5895} & \textbf{.4030} & \textbf{.5626} \\
                \midrule 
                S2L (ours)  & \xmark & Phenom1 & .4688 & .2729 & .2001 & .3139 & .5956 & .4374 & .3430 & .4587\\
                & & \& MolGPS &  &  &  &  &  &  &  &  \\
                S2L (ours)  & sigmoid & Phenom1 & .4983 & .3230 & .2397 & .3537 & .6343 & .5035 & .3790 & .5056\\
                & & \& MolGPS &  &  &  &  &  &  &  &  \\
                S2L (ours)  & logarithm & Phenom1 & .5101 & .3589 & .2535 & .3742 & .6398 & .5660 & .3992 & .5350\\
                & & \& MolGPS &  &  &  &  &  &  &  &  \\
                S2L (ours)  & one-hot &Phenom1 & \textbf{.5370} & \textbf{.3720} & \colorbox{babyblue}{\textbf{.2676}} & \colorbox{babyblue}{\textbf{.3922}} & \textbf{.6870} & \textbf{.5888} & \colorbox{babyblue}{\textbf{.4326}} & \colorbox{babyblue}{\textbf{.5695}} \\
                & & \& MolGPS &  &  &  &  &  &  &  &  \\
                \bottomrule
            \end{tabular}
            \end{adjustbox}
    \label{tab:appendix_cross_dose_results_all_mols}
\end{table}

\begin{table}
    \centering
  \caption{\textbf{Evaluation on held-out concentrations for active and inactive perturbations} Average Top-1\% and Top-5\% Recall accuracy of methods utilizing different contrastive learning methods. Best performing methods are highlighted in \textbf{bold}.}
  \begin{adjustbox}{max width=\textwidth}
    \begin{tabular}{ccccccccccc}
        \toprule
        &     & & \multicolumn{4}{c}{\textbf{top-1\%}} & \multicolumn{4}{c}{\textbf{top-5\%}} \\
        \cmidrule(lr){4-7} \cmidrule(lr){8-11}
         Loss & Explicit & Modality & Unseen Images & Unseen Images +& Unseen Dataset & Avg. & Unseen Images & Unseen Images +& Unseen Dataset & Avg. \\
        & Concentration &  &  & Unseen Molecules & (0-shot) & &  & Unseen Molecules & (0-shot) & \\
        \toprule
        CLIP &  \xmark & Phenom1  & .1684 & .1111 & .0964 & .1253 & .3916 & .2545 & .2356 & .2476 \\
        \midrule
        Hopfield-CLIP  & \xmark & Phenom1 & .1290 &	.0921 & .0756 & .0989 & .3485 & .2287 & .2095 & .2217 \\
        \midrule
        InfoLOOB & \xmark & Phenom1 & .1715 & .1114 & .0948 & .1259 & .3944 & .2578 & .2349 & .2495 \\
        \midrule
        CLOOME  & \xmark & Phenom1 &  .1745 & .1088 & .0910 & .1248 & .4093 & .2487 & .2355 & .2439 \\
        CLOOME  & sigmoid & Phenom1 &  \textbf{.2573}	& \textbf{.1208}	& \textbf{.1062} & \textbf{.1614} & \textbf{.5169} & \textbf{.2638}	& \textbf{.2513} & \textbf{.3440} \\
        CLOOME  & logarithm & Phenom1 &  .2379 & .1081	& .0992	& .1484 & .4958 & .2444	& .2324	& .3242 \\
        CLOOME  & one-hot & Phenom1 &  .2346 & .0970	& .0974	& .1430 & .5014 & .2224	& .2348	& .3195 \\
        \midrule
        DCL & \xmark & Phenom1 & .3516 & .1655 & .1533 & .2235 & .5693 & .3125 & .3006 & .3082 \\ 
        DCL & sigmoid & Phenom1 & \textbf{.4741} & \textbf{.1725} & \textbf{.1726} & \textbf{.2731} & \textbf{.6637} & \textbf{.3261} & \textbf{.3105} & \textbf{.3204} \\ 
        DCL & logarithm & Phenom1 & .4286 & .1596 & .1581 & .2488 & .6244 & .3071 & .3032 & .3056 \\ 
        DCL & one-hot & Phenom1 & .4308 & .1495 & .1600 & .2468 & .6244 & .2938 & .3015 & .2966 \\ 
        \midrule
        CWCL  & \xmark & Phenom1 & .4126 &  .1801 & .1667 & .2531 & .6128 & .3266 & .3066 & .3194\\
        CWCL  & sigmoid & Phenom1 & \textbf{.5112} & \textbf{.1856} & \textbf{.1811} & \textbf{.2926} & \textbf{.6901} & \textbf{.3384} & \textbf{.3190 } & \textbf{.3314} \\
        CWCL  & logarithm & Phenom1 & .4664 & .1696 & .1709 & .2690 & .6502 & .3195 & .3066 & .3148\\
        CWCL  & one-hot & Phenom1 & .4681 &	.1612 & .1734 & .2676 & .6465 & .3019 & .3104 & .3050\\
        \midrule
        SigLip & \xmark & Phenom1  & .3942 & .1578 & .1390 & .2303 & .5931 & .3015 & .2737 & .2914 \\
        SigLip & sigmoid & Phenom1  & \textbf{.5392} & \textbf{.1828} & \textbf{.1710} & \textbf{.2977} & \textbf{.7102} & \textbf{.3399} & \textbf{.3121} & \textbf{.3298} \\
        SigLip & logarithm & Phenom1  & .5022 & .1698 & .1669 & .2796 & .6841 & .3240 & .3068 & .3177 \\
        SigLip & one-hot & Phenom1  & .4657 & .1443 & .1451 & .2517 & .6544 & .2879 & .2790 & .2847 \\
        \midrule
        S2L (ours)  & \xmark & Phenom1  & .5336 & .1842 & .1713 & .2963 & .6961 & .3322 & .3045 & .3221 \\
        S2L (ours)  & sigmoid & Phenom1  & \colorbox{babyblue}{\textbf{.5409}} & \colorbox{babyblue}{\textbf{.1899}} & \textbf{.1753} & \textbf{.3020} & \colorbox{babyblue}{\textbf{.7178}} & \colorbox{babyblue}{\textbf{.3469}} & \textbf{.3201} & \textbf{.3372}\\
        S2L (ours)  & logarithm & Phenom1  & .5036 & .1791 & .1727 & .2851 & .6925 & .3342 & .3157 & .3275\\
        S2L (ours)  & one-hot & Phenom1  & .4726 & .1537 & .1521 & .2595 & .6696 & .2998 & .2887 & .2958\\
        \midrule 
        S2L (ours)  & \xmark & Phenom1 & .5248 & .1829 & .1910 & .2996 & .6904 & .3268 & .3305 & .3281 \\
        & & \& MolGPS &  &  &  &  &  &  &  &  \\
        S2L (ours)  & sigmoid & Phenom1 & \textbf{.5338} & \textbf{.1897} & \colorbox{babyblue}{\textbf{.2029}} & \colorbox{babyblue}{\textbf{.3088}} & \textbf{.7098} & \textbf{.3427} & \colorbox{babyblue}{\textbf{.3495}} & \colorbox{babyblue}{\textbf{.3452}} \\
        & & \& MolGPS &  &  &  &  &  &  &  &  \\
        S2L (ours)  & logarithm & Phenom1 & .4900 & .1839 & .2031 & .2923 & .6776 & .3354 & .3511 & .3411 \\
        & & \& MolGPS &  &  &  &  &  &  &  &  \\
        S2L (ours)  & one-hot & Phenom1 & .4622 & .1569 & .1762 & .2651 & .6578 & .3030 & .3187 & .3087 \\
        & & \& MolGPS &  &  &  &  &  &  &  &  \\
        \bottomrule
    \end{tabular}
    \end{adjustbox}
    \label{tab:appendix_whitin_dose_results_all_mols}
\end{table}

\subsection{Ablation Studies}\label{sc:ablation}

\begin{figure}[t]
    \centering
    \includegraphics[width=.9\textwidth]{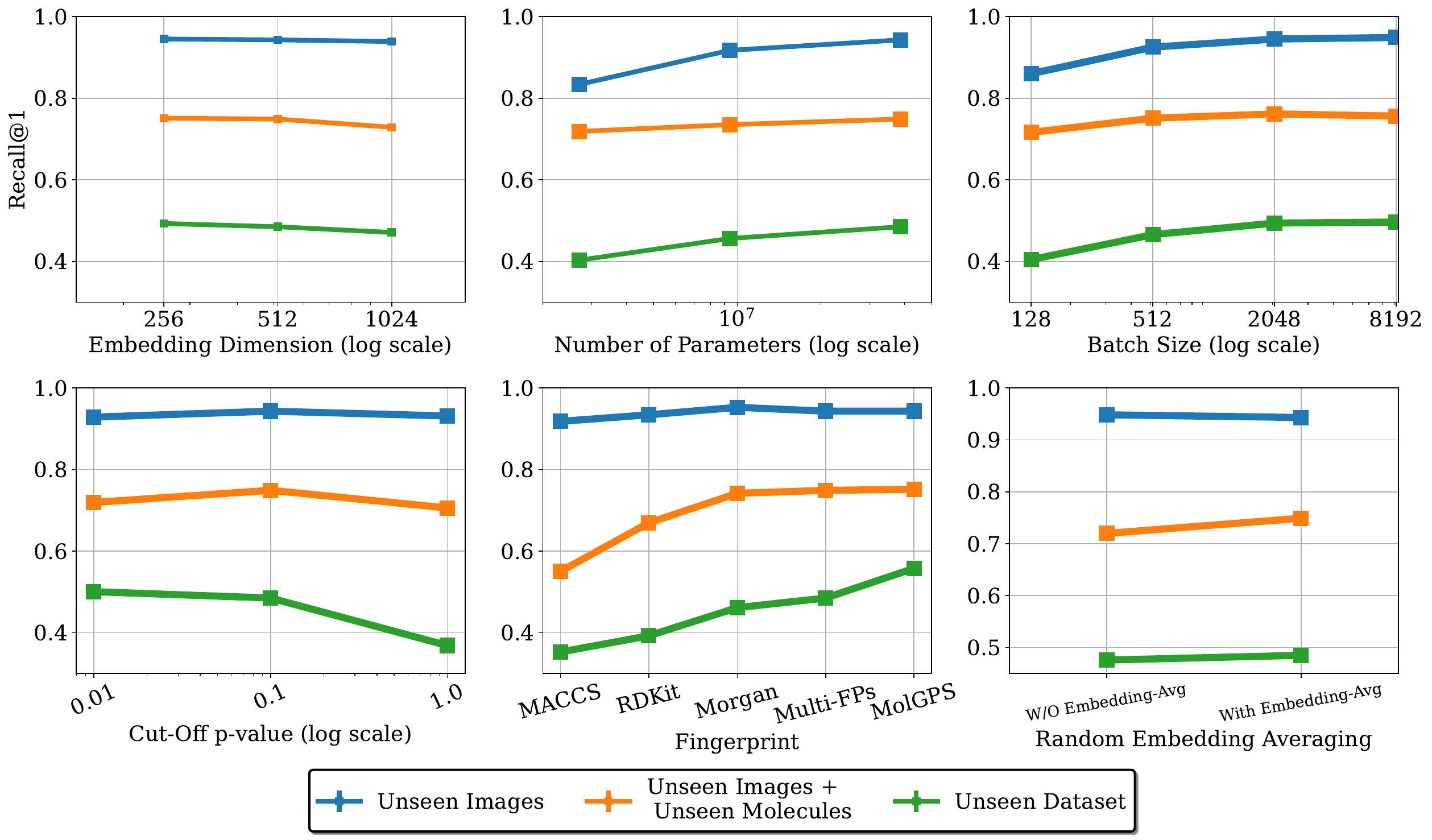}
    \caption{Ablations of top-1 \% recall accuracy with \textbf{(top-left)} the size of embedding dimension, \textbf{(top-center)} number of parameters, \textbf{(top-right)} batch size, \textbf{(bottom-left)} cutoff $p$ value, \textbf{(bottom-center)} fingerprint type, and \textbf{(bottom-right)} random batch averaging. Compact embedding sizes from pretrained models, larger number of parameters, larger batch sizes, lower cutoff p-values, pretrained MolGPS fingerprints and presence of random batch averagin improving retrieval of our MolPhenix framework.}
    \label{fig:ablations_full}
\end{figure}

Figure \ref{fig:ablations_full} and Table \ref{tab:ablation_batch_size}, \ref{tab:ablation_dim_size}, \ref{tab:ablation_pvalue}, \ref{tab:ablation_fp} and \ref{tab:ablation_REA} present top-1\% recall accuracy across for the full ablation study on the variation of MolPhenix key components.
We note that compact embedding sizes from pretrained models stabilize training. This indicates that embeddings are expressive and accurately capture intricate aspects of molecules. Larger batch sizes result in a greater number of negative samples, hence improving performance. This is in line with prior contrastive learning methods continuing to improve by increasing the batch size \cite{simclr2}. Increasing the number of parameters leads to more expressive models thereby enhancing retrieval performance. This result is in accordance with recent advances in language modelling and scaling laws across different data and compute budgets \cite{scalinglaws}.

\begin{table}[h]
    \centering
    \begin{adjustbox}{max width=.8\textwidth}
    \begin{tabular}{cccccc}
         \hline
          Model size & Depth & Width & Unseen images & Unseen images + & Unseen dataset\\
         & & & & Unseen molecules & (0-shot) \\
         \hline
         Tiny - 2.7m & 4 ResBlocks & 256 & .8337 & .7186 & .4030  \\
         Small - 9.4m & 6 ResBlocks & 512 & .9174 &  .7352 & .4562 \\
         Medium - 38.7m & 8 ResBlocks & 1024 & .9430 & .7490 &  .485 \\
         \hline
    \end{tabular}
    \end{adjustbox}
    \vspace{.1cm}
    \caption{Ablations across different model sizes. Larger capacity models are found to be more expressive.}
    \label{tab:ablation_model_size}
\end{table}

\begin{table}[h]
    \centering
    \begin{adjustbox}{max width=.6\textwidth}
    \begin{tabular}{cccc}
         \hline
          Batch size & Unseen images & Unseen images + & Unseen dataset\\
         & & Unseen molecules & (0-shot) \\
         \hline
         128 & .8600 & .7163 &  .4044  \\
         512 & .9252 & .7511 & .4657  \\
         2048 & .9450 & .7616 & .4940  \\
         8192 & .9489 & .7563 & .4966  \\
         \hline
    \end{tabular}
    \end{adjustbox}
    \vspace{.1cm}
    \caption{Ablation across different batch sizes. Larger batch sizes benefit contrastive learning.}
    \label{tab:ablation_batch_size}
\end{table}

\begin{table}[h]
    \centering
    \begin{adjustbox}{max width=.5\textwidth}
    \begin{tabular}{|c|c|c|c|}
         \hline
         Dim size & Unseen images & Unseen images + & Unseen dataset\\
         & & Unseen molecules & (0-shot) \\
         \hline
         256 & .9452 & .7510 & .4929  \\
         512 & .9430 & .7490 & .4850 \\
         1024 & .9392 & .7288 & .4710 \\
         \hline
    \end{tabular}
    \end{adjustbox}
    \caption{Ablation across different embedding dimensions. Compact embedding sizes capture more molecular information.}
    \label{tab:ablation_dim_size}
\end{table}

\begin{table}[h]
    \centering
    \begin{adjustbox}{max width=.5\textwidth}
    \begin{tabular}{|c|c|c|c|}
         \hline
         cut-off & Unseen images & Unseen images + & Unseen dataset\\
         & & Unseen molecules & (0-shot) \\
         \hline
         p < 1.0 & .9312 & .7057 & .3686  \\
         p < .1 & .9430 & .7490 & .4850 \\
         p < .01 & .9284 & .7192 & .5005 \\
         \hline
    \end{tabular}
    \end{adjustbox}
    \caption{Ablation across different p-value cutoff threhsolds. p values < .1 benefit retrieval of active molecules.}
    \label{tab:ablation_pvalue}
\end{table}

\begin{table}[h]
    \centering
    \begin{adjustbox}{max width=.5\textwidth}
    \begin{tabular}{|c|c|c|c|}
         \hline
         fingerprint & unseen images & unseen images + & unseen dataset\\
         & & unseen molecule &  \\
        \hline
         MACCS & .9180 & .5503 & .3526 \\
         RDKit &  .9341 & .6693 & .3925 \\
         Morgan & .9524 & .7417 & .4613 \\
         Multi-FPs & .9430 & .7490 & .485 \\
        Phenom1 + MolGPS & .9430 & .7514 & .5577 \\
         \hline
    \end{tabular}
    \end{adjustbox}
    \caption{Ablation across different fingerprint types. A combination of embeddings bootstrapped from Phenom1 and MolGPS significantly benefit retrieval.}
    \label{tab:ablation_fp}
\end{table}

\begin{table}[h]
    \centering
    \begin{adjustbox}{max width=.5\textwidth}
    \begin{tabular}{|c|c|c|c|}
         \hline
          & Unseen images & Unseen images + & Unseen dataset\\
         & & Unseen molecules & (0-shot) \\
         \hline
         W/O Random Embedding Avg. & .9482 & .7198 & .4759 \\
         With Random Embedding Avg. & .9430 & .7490 & .485 \\
         \hline
    \end{tabular}
    \end{adjustbox}
    \caption{Ablation across random embedding averaging. Utilizing random batch averaging stabilizes training and benefits retrieval.}
    \label{tab:ablation_REA}
\end{table}

\subsection{Investigating Other Pre-trained Phenomic Encoders}
To investigate the impact of pre-trained encoders, we perform additional experiments evaluating a supervised phenomic image encoder (Table~\ref{tab:AdaBN}). Instead of Phenom1, we trained Molphenix framework using AdaBN, a CNN-based supervised phenomic encoder, with an analogous implementation discussed in~\cite{rxrx1}. We find that the general trends between Phenom1 and AdaBN are consistent with a slight decrease in overall performance. These findings provide additional support to the generality of the proposed guidelines.

\begin{table}[h]
    \centering
     \caption{Evaluation on \textbf{cumulative concentrations} while using \textbf{AdaBN}. Molphenix is trained on combination of RDKIT, MACCS, and Morgan fingerprints in this experiment}
  \begin{adjustbox}{max width=\textwidth}
    \begin{tabular}{ccccccccccc}
    % \toprule
        \toprule
        Method  & Explicit  & Modality & Unseen Images & Unseen Images +& Unseen Dataset & Avg. & Unseen Images & Unseen Images +& Unseen Dataset & Avg. \\
         & Concentration &  &  & Unseen Molecules & (0-shot) & &  & Unseen Molecules & (0-shot) & \\
         \toprule
        &   &  & \multicolumn{4}{c}{\textbf{top-1\% active molecules}} & \multicolumn{4}{c}{\textbf{top-5\% active molecules}} \\
         \cmidrule(lr){4-7} \cmidrule(lr){8-11}

        MolPhenix &- &AdaBN &.8568 &.5336 &.3525 &.581 &.9562 &.7603 &.5772 &.7646 \\
        MolPhenix &sigmoid &AdaBN &.911 &.5858 &.4 &.6323 &.971 &.7997 &.6203 &.797 \\
        MolPhenix &logarithm &AdaBN &.9155 &.6106 &.4242 &.6501 &.9729 &.8332 &.6503 &.8188 \\
        MolPhenix &one-hot &AdaBN &.9187 &.6125 &.4225 &.6512 &.9744 &.8302 &.6419 &.8155 \\

        \toprule
%     \end{tabular}
%     \end{adjustbox}
%     \label{tab:appendix_cross_dose_results}
% \end{table}

% \begin{table}
%     \centering
%      \caption{Evaluation on \textbf{cumulative concentrations} for all molecules while using \textbf{AdaBN}}
%   \begin{adjustbox}{max width=\textwidth}
%     \begin{tabular}{ccccccccccc}
%     % \toprule
%         \toprule
        &   &  & \multicolumn{4}{c}{\textbf{top-1\% all molecules}} & \multicolumn{4}{c}{\textbf{top-5\% all molecules}} \\
         \cmidrule(lr){4-7} \cmidrule(lr){8-11}
%         Method  & Explicit  & Modality & Unseen Images & Unseen Images +& Unseen Dataset & Avg. & Unseen Images & Unseen Images +& Unseen Dataset & Avg. \\
%          & Concentration &  &  & Unseen Molecules & (0-shot) & &  & Unseen Molecules & (0-shot) & \\
        % \toprule
        MolPhenix &- &AdaBN &.4593 &.2409 &.1599 &.2867 &.5983 &.4081 &.285 &.4305 \\
        MolPhenix &sigmoid &AdaBN &.5104 &.3142 &.1957 &.3401 &.6496 &.5165 &.331 &.499 \\
        MolPhenix &logarithm &AdaBN &.5379 &.3393 &.2071 &.3614 &.6867 &.5561 &.3606 &.5345 \\
        MolPhenix &one-hot &AdaBN &.5476 &.3425 &.2082 &.3661 &.7007 &.5641 &.3603 &.5417 \\

        \bottomrule
    \end{tabular}
    \end{adjustbox}
    \label{tab:AdaBN}
\end{table}

\subsection{Integrating MolGPS Embeddings With Other Fingerprints}

Molphenix architecture is flexible, allowing that the proposed components be replaced by other phenomic or molecular pretrained models. We leveraged from MolGPS, which is a MPNN based GNN model with 1B parameters which allows us to maximize architecture expressivity while minimizing the risk of overfitting~\cite{gps++, molgps}. For additional investigation, we combine MolGPS molecular embeddings with RDKIT, MACCS, and Morgan fingerprints and show that they can provide Molphenix with richer molecular information and yields overall higher performance of MolPhenix in both cumulative and held-out concentration scenarios. Results for active and all molecules retrieval of Molphenix trained on the discussed combinational molecular embeddings are available in table~\ref{tab:integrating_cumilative} and \ref{tab:integrating_heldout}.

\begin{table}[h]
    \centering
     \caption{Evaluation on \textbf{cumulative concentrations} while \textbf{combining MolGPS, RDKIT, MACCS, and Morgan fingerprints}.}
  \begin{adjustbox}{max width=\textwidth}
    \begin{tabular}{ccccccccccc}
    % \toprule
        \toprule
        Method  & Explicit  & Modality & Unseen Images & Unseen Images +& Unseen Dataset & Avg. & Unseen Images & Unseen Images +& Unseen Dataset & Avg. \\
         & Concentration  &  &  & Unseen Molecules & (0-shot) & &  & Unseen Molecules & (0-shot) & \\
         \toprule
         &   &  & \multicolumn{4}{c}{\textbf{top-1\%  active molecules}} & \multicolumn{4}{c}{\textbf{top-5\%  active molecules}} \\
         \cmidrule(lr){4-7} \cmidrule(lr){8-11}
        % \toprule
        MolPhenix &- &Phenom1 \& MolGPS \& 3 fps &.9185 &.7212 &.4717 &.7038 &.9784 &.8805 &.718 &.859 \\
        MolPhenix &sigmoid & Phenom1 \& MolGPS \& 3 fps &.9395 &.7408 &.5119 &.7307 &.9817 &.8932 &.7458 &.8736 \\
        MolPhenix &logarithm & Phenom1 \& MolGPS \& 3 fps &.9454 &.7798 &.5658 &.7637 &.9815 &.9163 &.7849 &.8942 \\
        MolPhenix &one-hot &Phenom1 \& MolGPS \& 3 fps &.9419 &.7687 &.5526 &.7544 &.9807 &.9113 &.7681 &.8867 \\

        \toprule
%     \end{tabular}
%     \end{adjustbox}
%     \label{tab:appendix_cross_dose_results}
% \end{table}

% \begin{table}
%     \centering
%      \caption{Evaluation on \textbf{cumulative concentrations} for \textbf{all} molecules while combining MolGPS \& other fps.}
%   \begin{adjustbox}{max width=\textwidth}
%     \begin{tabular}{ccccccccccc}
%     % \toprule
%         \toprule
        &   &  & \multicolumn{4}{c}{\textbf{top-1\%  all molecules}} & \multicolumn{4}{c}{\textbf{top-5\%  all molecules}} \\
         \cmidrule(lr){4-7} \cmidrule(lr){8-11}
%         Method  & Explicit  & Modality & Unseen Images & Unseen Images +& Unseen Dataset & Avg. & Unseen Images & Unseen Images +& Unseen Dataset & Avg. \\
%          & Concentration  &  &  & Unseen Molecules & (0-shot) & &  & Unseen Molecules & (0-shot) & \\
        % \toprule
        MolPhenix &- &Phenom1 \& MolGPS \& 3 fps &.4764 &.3011 &.2068 &.3281 &.604 &.4647 &.3415 &.4701 \\
        MolPhenix &sigmoid &Phenom1 \& MolGPS \& 3 fps &.5076 &.342 &.2382 &.3626 &.6383 &.521 &.3769 &.512 \\
        MolPhenix &logarithm &Phenom1 \& MolGPS \& 3 fps &.525 &.379 &.2648 &.3896 &.658 &.5743 &.411 &.5478 \\
        MolPhenix &one-hot &Phenom1 \& MolGPS \& 3 fps &.5355 &.3845 &.265 &.395 &.6862 &.5916 &.4233 &.567 \\

        \bottomrule
    \end{tabular}
    \end{adjustbox}
    \label{tab:integrating_cumilative}
\end{table}
% \vspace{-.5cm}

\begin{table}[h]
    \centering
     \caption{Evaluation on \textbf{heldout concentrations} while \textbf{combining MolGPS, RDKIT, MACCS, and Morgan fingerprints}.}
  \begin{adjustbox}{max width=\textwidth}
    \begin{tabular}{ccccccccccc}
    % \toprule
        \toprule
        Method  & Explicit  & Modality & Unseen Images & Unseen Images +& Unseen Dataset & Avg. & Unseen Images & Unseen Images +& Unseen Dataset & Avg. \\
         & Concentration  &  &  & Unseen Molecules & (0-shot) & &  & Unseen Molecules & (0-shot) & \\
        \toprule
         &   &  & \multicolumn{4}{c}{\textbf{top-1\%  active molecules}} & \multicolumn{4}{c}{\textbf{top-5\%  active molecules}} \\
         \cmidrule(lr){4-7} \cmidrule(lr){8-11}
        MolPhenix &- &Phenom1 \& MolGPS \& 3 fps &.8364 &.5115 &.4263 &.5914 &.9640 &.7363 &.6850 &.7951 \\
        MolPhenix &sigmoid &Phenom1 \& MolGPS \& 3 fps &.8300 &.5021 &.4363 &.5895 &.9640 &.7409 &.6931 &.7993 \\
        MolPhenix &logarithm &Phenom1 \& MolGPS \& 3 fps &.8112 &.5107 &.4376 &.5865 &.9544 &.7406 &.6866 &.7939 \\
        MolPhenix &one-hot &Phenom1 \& MolGPS \& 3 fps &.7467 &.4409 &.3830 &.5235 &.9320 &.6827 &.6520 &.7556 \\

%         \bottomrule
%     \end{tabular}
%     \end{adjustbox}
%     \label{tab:appendix_cross_dose_results}
% \end{table}

% \begin{table}
%     \centering
%      \caption{Evaluation on \textbf{heldout concentrations} for \textbf{all} molecules while combining MolGPS \& other fps.}
%   \begin{adjustbox}{max width=\textwidth}
%     \begin{tabular}{ccccccccccc}
%     % \toprule
        \toprule
        &   &  & \multicolumn{4}{c}{\textbf{top-1\%  all molecules}} & \multicolumn{4}{c}{\textbf{top-5\%  all molecules}} \\
         \cmidrule(lr){4-7} \cmidrule(lr){8-11}
%         Method  & Explicit  & Modality & Unseen Images & Unseen Images +& Unseen Dataset & Avg. & Unseen Images & Unseen Images +& Unseen Dataset & Avg. \\
%          & Concentration  &  &  & Unseen Molecules & (0-shot) & &  & Unseen Molecules & (0-shot) & \\
        % \toprule
        MolPhenix &- &Phenom1 \& MolGPS \& 3 fps &.5339 &.1980 &.1966 &.3095 &.6968 &.2909 &.4274 &.4717 \\
        MolPhenix &sigmoid &Phenom1 \& MolGPS \& 3 fps &.5463 &.2026 &.2066 &.3185 &.7179 &.3116 &.4359 &.4885 \\
        MolPhenix &logarithm &Phenom1 \& MolGPS \& 3 fps &.5247 &.2009 &.2078 &.3111 &.7067 &.3133 &.4319 &.4840 \\
        MolPhenix &one-hot &Phenom1 \& MolGPS \& 3 fps &.4690 &.1653 &.1756 &.2700 &.6635 &.2592 &.4118 &.4448 \\

        \bottomrule
    \end{tabular}
    \end{adjustbox}
    \label{tab:integrating_heldout}
\end{table}